\documentclass[showpacs,aps,floatfix,prd,preprint,superscriptaddress]{revtex4-1}
\usepackage{xfrac}
\usepackage{amssymb}
\usepackage{braket}
\usepackage{graphicx}
\usepackage{verbatim}
\usepackage{etoolbox}
\usepackage{amsfonts}
\usepackage{amsmath}
\usepackage[utf8]{inputenc}
\usepackage{mathtools}
\usepackage{soul,xcolor,colortbl}
\usepackage{hyperref} \hypersetup{colorlinks=true,citecolor=blue,linkcolor=blue,urlcolor=blue}
\makeatletter \renewcommand\frontmatter@abstractwidth{\dimexpr\textwidth\relax} \makeatother

\usepackage{multirow}
\usepackage{longtable}
\usepackage{cellspace}
\usepackage{resizegather}
\usepackage{dcolumn}
\usepackage[normalem]{ulem}

\usepackage{bm}
\usepackage{array}
\newcolumntype{C}[1]{>{\centering\arraybackslash}p{#1}}\usepackage{soul}
\definecolor{Gray}{gray}{0.85}

\usepackage{color}
\usepackage{morefloats}

\definecolor{Gray}{gray}{0.9}
\definecolor{LightCyan}{rgb}{0.88,1,1}
\definecolor{green}{rgb}{0.5451,0.2706,0.0745}

\begin{document}

\title{Vibrational fingerprintings for chemical recognition of biominerals}

\author{Arrigo Calzolari}
\email {Email: arrigo.calzolari@nano.cnr.it}
\affiliation{CNR-NANO, Istituto Nanoscienze, Centro S3, via Campi 213A, I-41125 Modena, IT}
\affiliation{Department of Physics, University of North Texas, Denton, TX 76203, USA}

\author{Barbara Pavan}
\affiliation{Department of Chemistry and Science of Advanced Materials Program, Central Michigan University, Mt. Pleasant, MI 48859 USA}

\author{Stefano Curtarolo}
\email {Email: stefano@duke.edu}
\affiliation{Materials Science, Electrical Engineering, Physics and Chemistry, Duke University, Durham NC, 27708 USA}
\affiliation{Center for Materials Genomics, Duke University, Durham, NC 27708, USA}

\author{Marco Buongiorno Nardelli}
\affiliation{Department of Physics, University of North Texas, Denton, TX 76203, USA}
\affiliation{Center for Materials Genomics, Duke University, Durham, NC 27708, USA}

\author{Marco Fornari}
\affiliation{Department of Chemistry and Science of Advanced Materials Program, Central Michigan University, Mt. Pleasant, MI 48859 USA}
\affiliation{Center for Materials Genomics, Duke University, Durham, NC 27708, USA}

\date{\today}

\begin{abstract}
Pathologies associated with calcified tissue, such as osteoporosis, demand
{\it in vivo} and/or {\it in situ} spectroscopic analysis to assess the role of chemical
substitutions  in the inorganic component. High energy X-ray or NMR spectroscopies are often impractical
or damaging in biomedical conditions. Low energy spectroscopies, such as IR and Raman techniques, are often the best alternative. In apatite biominerals, the vibrational
signatures of the phosphate group are generally used
as fingerprint of the materials although they provide only limited information. Here, we have used first principles calculations to
 unravel the complexity of the complete vibrational spectra of apatites.
We determined the spectroscopic features of all the phonon modes of fluor-apatite, hydroxy-apatite,
and carbonated fluoroapatite beyond the analysis of the phosphate groups, focusing on the effect of local corrections induced by the crystalline environment and the specific mineral composition.
This provides a clear  and unique reference to discriminate structural and chemical variations
in biominerals, opening the way to a widespread application of non-invasive spectroscopies for {\it in vivo} diagnostics, and biomedical
analysis.
\end{abstract}

\maketitle

\section*{INTRODUCTION}

Calcium orthophosphates form the main inorganic component of normal and pathological calcified tissues in vertebrates \cite{Dorozhkin2009}. 
Apatites (general formula Ca$_5$(PO$_4$)$_3$X  with X = F, Cl, OH), with various degrees of substitutions and defects, contribute ~60-70 $\%$ to human bones and are of great importance in widespread diseases such as cancer \cite{tampieri13}, osteoporosis \cite{Faibish2006,tampieri07} and craniosynostosis \cite{AlRekabi2017}.
Most of calcium orthophosphates are biocompatible and plays critical roles in biomineralzation processes. For instance, apatite bioceramics are used {in regenerative medicine} as bone repair and augmentation, as bone cements, as coatings for bio-inert prostheses, and for bone scaffolding. \cite{tampieri00,Rosin01,Daculsi04,LeGeros08}
Calcium orthophosphates are also commonly found in the Earth crust and are considered one of the most abundant sources of phosphorous in the marine environment \cite{Tribble95} and are relevant in the geophysics of pegmatite melts \cite{mona09}, especially for their ability to readily accept substitution(s) from a large variety of ions. For the same reason, they are attractive for nuclear waste management and water remediation \cite{zheng07,wellman08}. 
Very recently, calcium orthophosphates have been recognized also in plant leaves,
opening brand new scenarios on the role of carbonate biominerals in replacing the common plant biomineral silica \cite{Weigend16,Weigend18}.

The functional bio-properties of apatites rely heavily on the shape of the nanocrystals \cite{stifler18} which, in turn, depends on the specific composition. For instance, the fluorine rich apatite platelets in teeth are neddle-like \cite{stifler17}  whereas the spheroidal nano-crystals in bones are mainly carbonated hydroxy-apatites \cite{lotsari18}.
Therapeutic treatments for osteoporosis, such as fluorine therapy, aim to influence the chemical composition of the mineral component of bones in order to improve the mechanical properties that result from the morphology and the size of the apatite crystals dispersed in the collagen matrix \cite{Faibish2006,Nudelman10}. {More generally, clinical applications require standardized protocols for biomaterials recognition, that, to date,  have not been analytically defined, since the characterization of the physicochemical properties of biogenic minerals is far from being completed.}

Many experimental techniques have been used to inspect the subtle structural changes associated with the specific chemical composition in apatites. {X-ray spectroscopies (such as XPS photoemission \cite{Suchanek04,Elkabouss04,Ohtsu13,stifler18}  and NEXAFS absorption \cite{kruse09,nakata09,kato14} spectra)  or NMR \cite{pan02,panda03,hu10,pavan12} result to be very efficient in this regards, being able to relate single spectral features to both atomic structure and chemical composition. As illustrative example, we included in the Supplementary Information (SI) simulated the X-ray core level photoemission spectra  of the O$_{1s}$ level relative to three prototypical apatite structures, which differ in the chemical composition (see also Figure \ref{fig:1}). The resulting shifts of the core level binding energies clearly pinpoint to the different local chemical environment induced by the specific atomic substitution (e.g. fluorine with hydroxyl or a phosphate with a carbonate unit). Nonetheless, such analysis requires high-cost instrumentations usually available in dedicated facilities (e.g. synchrotron radiation lines) and works in conditions (e.g. ultra high vacuum, high-energy radiations, controlled temperature) that make their use unsuitable for clinical and biomedical applications. Low-energy vibrational techniques
(e.g. IR, and Raman) are often desired due to their non-invasive non-destructive character  which is suitable for {\it in vivo} and/or {\it in situ} analysis.
In those case, an overall map of the vibrational spectra is essential to fingerprint the structural and chemical effects of biominerals. Unfortunately, such structural and chemical features are much more difficult to identify in vibrational spectroscopies (compared to X-ray ones) as they cause complex modifications of the spectra  (e.g. degeneracy breaking,  structured multipeaks, and intensity changes of the active modes) that are not easy to disentangle. First principles calculations, having a direct access to structural,  electronic, and vibrational properties of materials, clarify the subtle interplay between chemical composition and optical spectra; in turn they helps to understand chemistry-morphology relationships in biominerals.}

So far, the standard approach to inspect apatites with vibrational spectroscopies was the recognition, within the entire spectra, of the four characteristic vibrational modes of the phosphate subunits.
In apatites, the phosphate groups are stabilized as orthophophate anions PO$_4^{3-}$  (i.e. fully deprotonated form), electrostatically compensated by the surrounding Ca$^{2+}$ cations.
However, due to its high reactivity, there are no vibrational spectra available in the literature for the isolated monophophate in the gas phase, and such measurements are usually performed in water solution.
Despite the profuse efforts \cite{schultze73, mason03,chapman64,preston79,cherif00, rudolph10,ajmal12,ajmal13}, the  characterization of the vibrational modes of orthophosphate anion in aqueous solution is {\it per se} a huge challenge, since PO$_4^{3-}$  is strongly hydrated and the ionic equilibrium with the intermediate
hydrogenated phases (HPO$_4^{2-}$ and H$_2$PO$_4^{-}$) is complicated.
This caused an unsolved variability
concerning the frequency position of the fundamental modes, the depolarization degree, and the appearance/disappearance of extra modes not due to the symmetry breaking in the  PO$_4^{3-}$ ion.
{Furthermore, since orthophophate anions are common to every apatite structure, it is  evident that the simple detection of the PO$_4^{3-}$ modes can not be sufficient to distinguish the composition
variety in the apatite class. }

In this paper, using {\it ab initio} theoretical methods, we  first investigate  the effects of the solvent on the vibrational properties of PO$_4^{3-}$, then we illustrate the dependence of the vibrational modes of the phosphate group in fluor-apatite, hydroxy-apatite, and carbonated fluor-apatite crystals. We also discuss features of the vibrational spectrum associated with crystalline effects and chemical composition. Our results point  clearly to the effect of the local environment on the overall vibrational properties of the system,
providing a set of representative vibrational markers univocally associated to microscopic nature of each crystal.
This provides a quantitative reference for experiments to improve the current understanding in investigative and therapeutic studies of calcified tissues. More generally, the systematic identification of such chemical fingerprints is a mandatory step towards the standardization of clinical protocols.

\section*{RESULTS AND DISCUSSION}

\subsection*{Phosphate's vibrational modes}

The  PO$_4^{3-}$ ion has a $T_d$ tetrahedral symmetry and nine vibrational modes corresponding to the representation
$\Gamma_{vib}= 2T+E+A$; the four different frequencies are labelled $\nu_1$, $\nu_2$, $\nu_3$, and $\nu_4$.
The non-degenerate  frequency $\nu_1$ corresponds to the  symmetric P-O stretching (A$_1$ symmetry), $\nu_2$ corresponds to the doubly degenerate O-P-O
bending mode (E symmetry); $\nu_3$ frequency arises from the triply degenerate T$_2$ mode involving the asymmetric P-O stretching and also P motion.
The $\nu_4$ mode (T$_2$ symmetry)  is associated to the triple degenerate asymmetric O-P-O bending.
All the modes are Raman active, while only $\nu_3$ and $\nu_4$ are infrared active.
When the orthophosphates are in aqueous solution to form HPO$_4^{2-}$ and H$_2$PO$_4^{-}$ ions the symmetry is reduced to $C_{3v}$  and $C_{2v}$, respectively.
This implies the progressive split of the degenerate vibrations, the shift of the non-degenerate ones and activation of the previously infrared-forbidden modes $\nu_1$ and $\nu_2$.
The coordination with the water molecules of the solution and the formation of H-bonds changes the charge distribution within the phosphate ions,  further affecting the vibrational spectrum of the PO$_4^{3-}$ ion \cite{tavan04,ebner05}.

The first step for the simulation of the (harmonic) vibrational properties is the identification of an equilibrium structure.
Due to the high charge unbalance, the ideal orthophosphate structure in vacuum is electronically not stable.
Thus, following experimental indications, we simulated the stability and the vibrational properties of PO$_4^{3-}$ ion in water solution.
Panel 1 of Figure \ref{}(a) shows the initial atomic configuration of the solvated  PO$_4^{3-}$ ion within the ESM. The simulation cell includes the orthophophate ion and a box of water molecules that entirely hosts the first hydration shell
 (16 H$_2$O molecules) \cite{ebner05}. The anion charge is neutralized by three Na$^+$ counterions \cite{preston79,tavan04}. During the total-energy-and-force optimization ({ i.e.} no thermal effects) the system undergoes a sequence of
protonation/deprotonation reactions that changes the final charge state of phosphate.
 Initially (panel 2), one water molecule close to the phosphate group spontaneously dissociates (H$_2$O $\rightarrow$ H$^+$+OH$^-$). Then (panel 3), the released proton is captured by PO$_4^{3-}$  that transforms  into HPO$_4^{2-}$.
 At the same time a second water molecule dissociates. Panels 4 and 5 show a progressive proton transfer that moves the remaining hydroxide molecule far away from the phosphate group
and closer to a Na$^+$ counterion. This relaxation process demonstrates that the fully deprotonated state is not stable at neutral pH, in agreement with the experiments
that adopted basic conditions (pH$>$13) \cite{schultze73,mason03} to stabilize  the orthophosphate anion.

\begin{figure}[!h!]
\begin{center}
\includegraphics[width=0.85\textwidth]{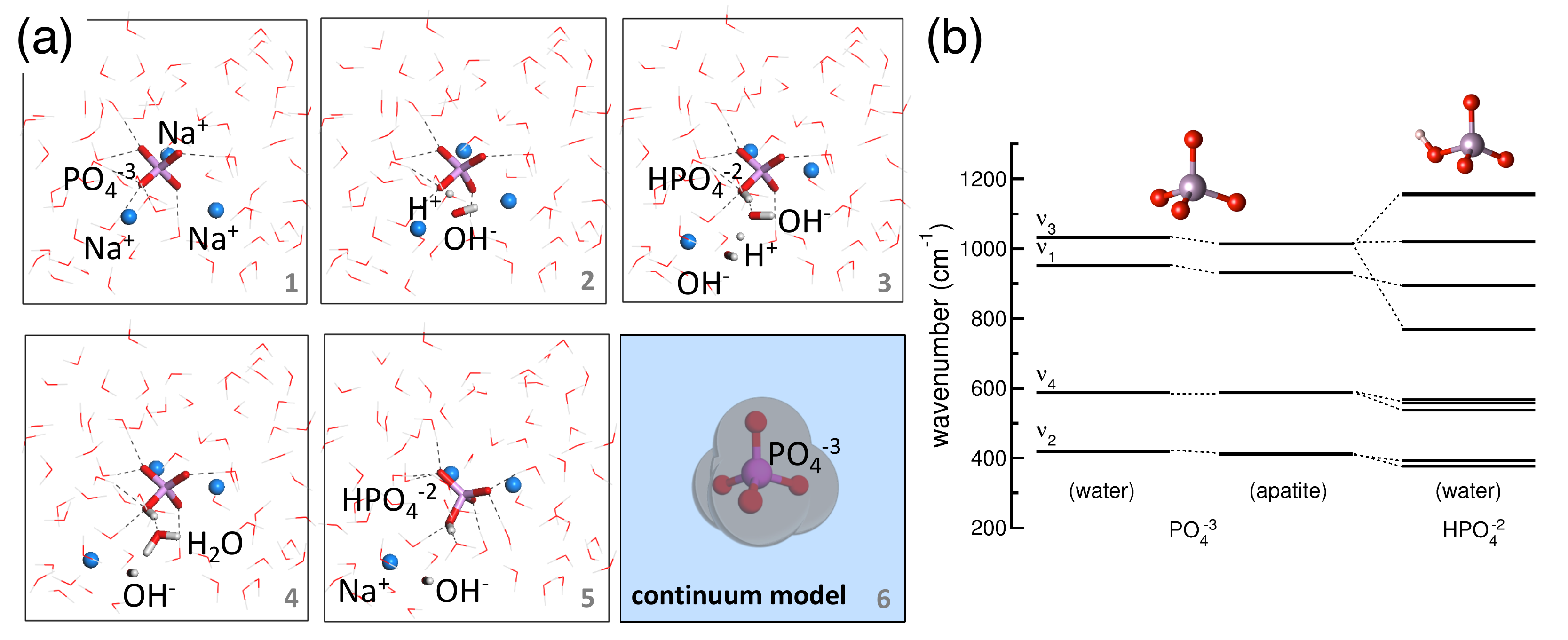}
  \caption{\small {\bf Solvated orthophophate.}
  (a) Atomic structure of solvated PO$_4^{3-}$ and HPO$_4^{2-}$ ions within the explicit solvent model (panels 1-5) and the continuum solvation model (panel 6). (b) Vibrational spectrum of PO$_4^{3-}$ and HPO$_4^{2-}$ ions embedded in continuum media reproducing either water or apatite
  dielectric properties.}
  \label{fig:2}
  \end{center}
\end{figure}

In order to overcome this problem, we optimized the orthophosphate ion within an implicit CSM which reproduces
the dielectric response of the solvent (panel 6). The vibrational frequencies of the corresponding ground state structure are  summarized in Figure\ref{fig:2}(b) and Table \ref{tab:po4}. Our results reproduce the four bands ($\nu_1-\nu_4$)
 of PO$_4^{3-}$, with the correct degeneracies derived by the symmetry analysis. The computed frequencies are in quantitative agreement with the experimental IR and Raman measurements \cite{schultze73,preston79}.
 In order to estimate the effect of the environment (i.e. water solvent { vs} apatite crystal) and of the hydration state we calculated the vibrational properties of PO$_4^{3-}$
 in a medium reproducing the dielectric properties of apatite ($\epsilon_r$=2.4) as well as
 the vibrational modes of HPO$_4^{2-}$ in aqueous solvent. The modification of the embedding polarization medium does not substantially affect the vibrational modes of orthophosphate, allowing us a direct comparison with apatite structures (see
 below), while the change of charge state ({ i.e.} hydration) imparts the splitting and shifting of the original vibrational bands expected by the symmetry analysis discussed above.
 Notably, in the case of HPO$_4^{2-}$ one negative frequency, associated to OH displacement, results from dynamical matrix diagonalization.
 We  accurately checked that the negative frequency is not an artefact or a limit of the simulation process, but rather it is the
 expression of a structural instability, associated to a saddle point of the total-energy surface.
 In other terms, this reproduces the tendency of the OH termination of moving in space,
being dynamically polarized by the environment ({ e.g.} the surrounding water molecules). A similar behavior has been reported in the case of the theoretical analysis of hydroxyapatite \cite{corno06}.

\subsection*{Structural features of apatites}

Apatite is a family of molecular crystals [M$_5$(PO$_4$)$_3$X]$_2$  that  occur in nature in several crystalline phases ({ e.g.} hexagonal, monoclinic) and
several  compositions (M=Ca$^{2+}$, Sr$^{2+}$; X=OH$^-$, F$^-$, Cl$^-$, Br$^-$). Isolated orthophosphate  groups (PO$_4^{3-}$) symmetrically arranged with respect to a mirror glide plane, corresponding to an ideal P6$_3$/m space group (n. 176, see Figure S1 in SI), are a typical feature of apatite crystals. Most of the experimental and theoretical spectroscopic investigations on apatite focused mostly on the phosphate group. The other elements that compose the mineral or the presence of
substituents and impurities received limited attention although their are essential to distinguish  different apatites.

\begin{figure}[!h!]
\begin{center}
\includegraphics[width=0.9\textwidth]{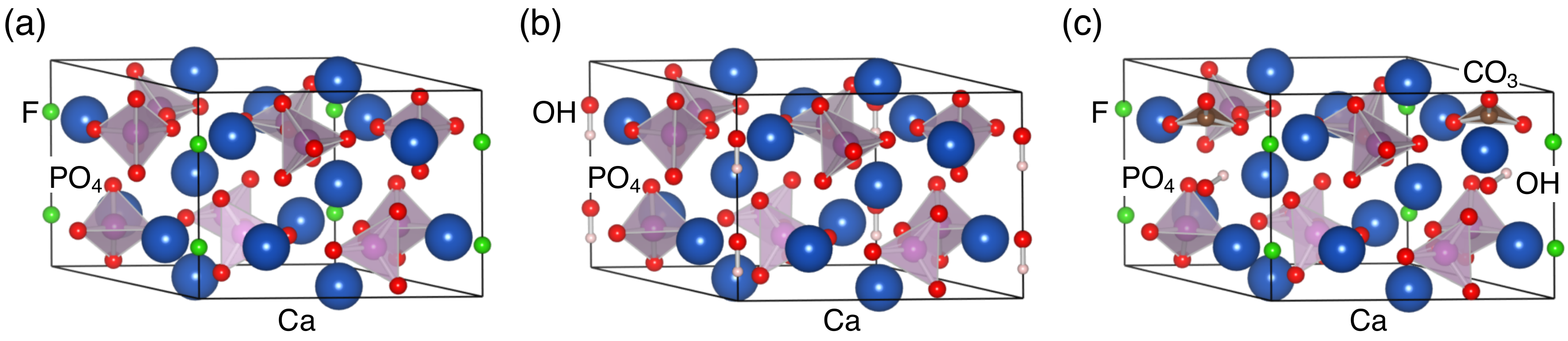}
  \caption{\small {\bf Crystal structures.}
  3D view of hexagonal crystal cells of (a)  fluoroapatite (FAp), (b) hydroxyapatite (HAp) and (c) carbonate-fluoroapatite (CFAp).}
  \label{fig:1}
  \end{center}
\end{figure}

In order to highlight subtleties beyond the phosphate group, we focus on three calcium structures, namely fluor-apatite (FAp), hydroxy-apatite (HAP), and carbonated fluor-apatite (CFAp) in their hexagonal crystalline phase.
The three crystal structures are shown in Figure \ref{fig:1}. Each system includes two [Ca$_5$(PO$_4$)$_3$X] (X=F,OH) units
symmetrically distributed on two parallel planes (labeled {\em a}-plane in Figure S1 of SI). Two inequivalent sets of Ca atoms forms either a columnar Ca-wires parallel to {\em c}-axis, or
or two triangles lying on two planes perpendicular to the {\em c}-axis, which form Ca-channels that host the non-metal X species (Figure S1).
Florine atoms (FAp, CFAp) also lie on the  {\em a}-plane, while OH$^-$ ions in HAp are aligned head-to-tail with the {\em c}-axis.
In the case of CFAp, one phosphate unit is substituted by a carbonate (CO$_3^{2-}$) group plus one OH$^-$ (Figure \ref{fig:1}c) to maintain the charge neutrality of the crystal.
Ground-state atomic relaxation has been reached without imposing any symmetry constrain. A summary of the resulting ground-state electronic properties is reported in SI.
Lattice parameters and internal coordinates have been fully optimized, the resulting lattice constants are a$_0$=9.51 \AA, c= 6.98 \AA;
a$_0$=9.42 \AA, c= 6.88 \AA~and a$_0$=9.52 \AA, c= 6.92 \AA~for FAp, HAp and CFAp, respectively,
in excellent agreement with experimental (Neutron scattering and X-ray) results \cite{saenger92,fleet95} and previous DFT calculations \cite{Rulis04,corno06}.
The optimized FAp geometry is very close to the experimental one \cite{sudarsan72} and converges to a highly symmetric structure associated to the $S_6 (-3)$ point group.
The inclusion of hydroxyl and carbonate units increases the atomic disorder within the unit cell and distorts the ideal tetrahedrical ($T_d$ symmetry) of the phosphate group, giving as a results
an overall reduction of the symmetry of the HAp and CFAp crystals.
In particular, the aligned hydroxyl groups in HAp breaks the mirror operation with respect to the glide plane, reducing the symmetry to $C_6 (6)$.
The inclusion of the CO$_3$ and OH fragments in CFAp  cancels all symmetry operations except identity ($C_1$ point group).

\subsection*{Orthophosphate's vibrations in apatites}

The phosphate group occurs in all apatites as fully protonated tetrahedral anion, the presence of
the other chemical species in the mineral such as Ca, F, OH, or CO$_3$, determines the symmetry of the
whole crystal and influences the functional properties. FAp has 42 atoms per cell corresponding to 126 phonon modes that can be classified according to the $S_6 (-3)$
symmetry as: A$_g$, A$_u$, $^1$E$_g$, $^1$E$_u$, $^2$E$_g$, and $^2$E$_u$;
acoustic modes are given by the representation $\Gamma_{ac}$=A$_u$+$^1$E$_u$+$^2$E$_u$, the remaining
{\it ungerade} modes are IR active and the {\it gerade} modes are Raman active. HAp has 132 independent phonon modes
classified by symmetry as: A, B, $^1$E$_1$, $^1$E$_2$, $^2$E$_1$, $^2$E$_2$.
Representation $\Gamma_{ac}$=A$_u$+$^1$E$_1$+$^2$E$_1$
describes the acoustic modes. Modes A, $^1$E$_1$, $^2$E$_1$ are IR active,  all modes except B are Raman active.
Finally, CFAp has 43 atoms per cell and 129 A phonon modes ($C_1$ symmetry), three of which are acoustic, all the others are IR and Raman active.

While most previous experimental and theoretical reports restrict their study to the Brillouin zone center ($\Gamma$) phonons, we provided a full characterization
of the  dielectric ({ e.g.} dielectric tensor, Born charges, Raman tensor) \cite{calzolari13} and vibrational  ({ e.g.} phonon dispersion and phonon density of states) properties of the selected FAp, HAp, and CFAp apatites. Results are summarized in SI. As observed for HPO$_4^{2-}$ case, one negative frequency constantly appears when OH
fragments are included in the crystal (namely, HAp and CFAp), which is an indication of the instability of OH bond orientation  due to the polar environment \cite{corno06}.
The phonon branches of all systems  (Figure S4, SI) are characterized by five main manifolds: one low energy branch
 ($<400$ cm$^{-1}$) involves contributions from all chemical species, and four bands in the range $[400-1300]$ cm$^{-1}$
 associated to the stretching  and bending vibrations of phosphate group. Carbonate group has characteristics optical modes at 818 and 1533 and 1553 cm$^{-1}$.
 Hydroxy substituents in HAp and CFAp have also characteristics optical modes in the frequency range $[700-800]$ cm$^{-1}$.  In the case of HAp, and CFAp, high
frequency  modes ($\sim$ 3800 cm$^{-1}$) associated to asymmetric OH stretching are also detected. The latter OH frequencies are blue-shifted
with respect to the experimental values ($\sim$ 3570 cm$^{-1}$) \cite{tsuda94,rehman97,cusco98}, since anharmonic corrections are not included in the simulations.

\begin{figure}[!h!]
\begin{center}
\includegraphics[width=0.45\textwidth]{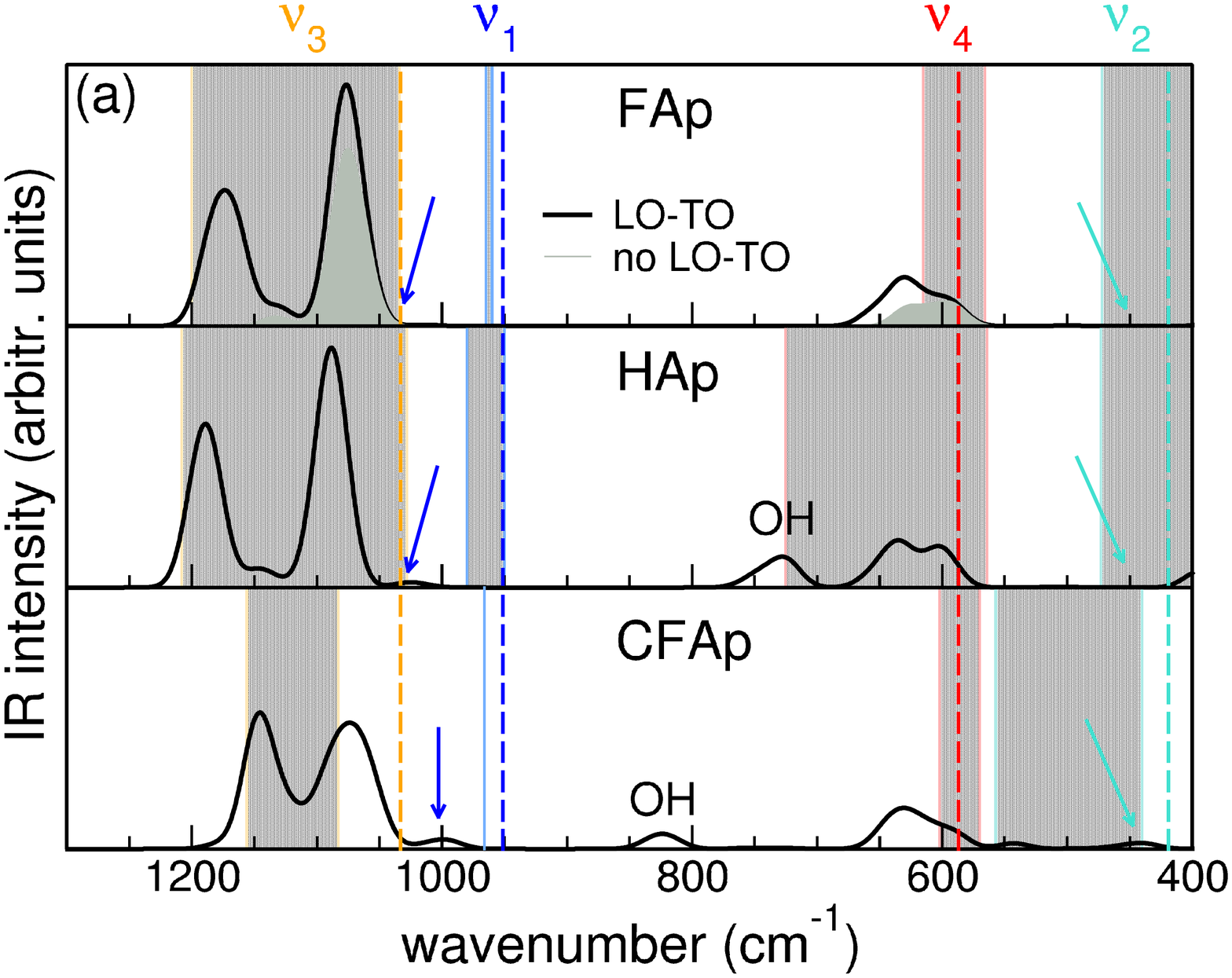}
\includegraphics[width=0.45\textwidth]{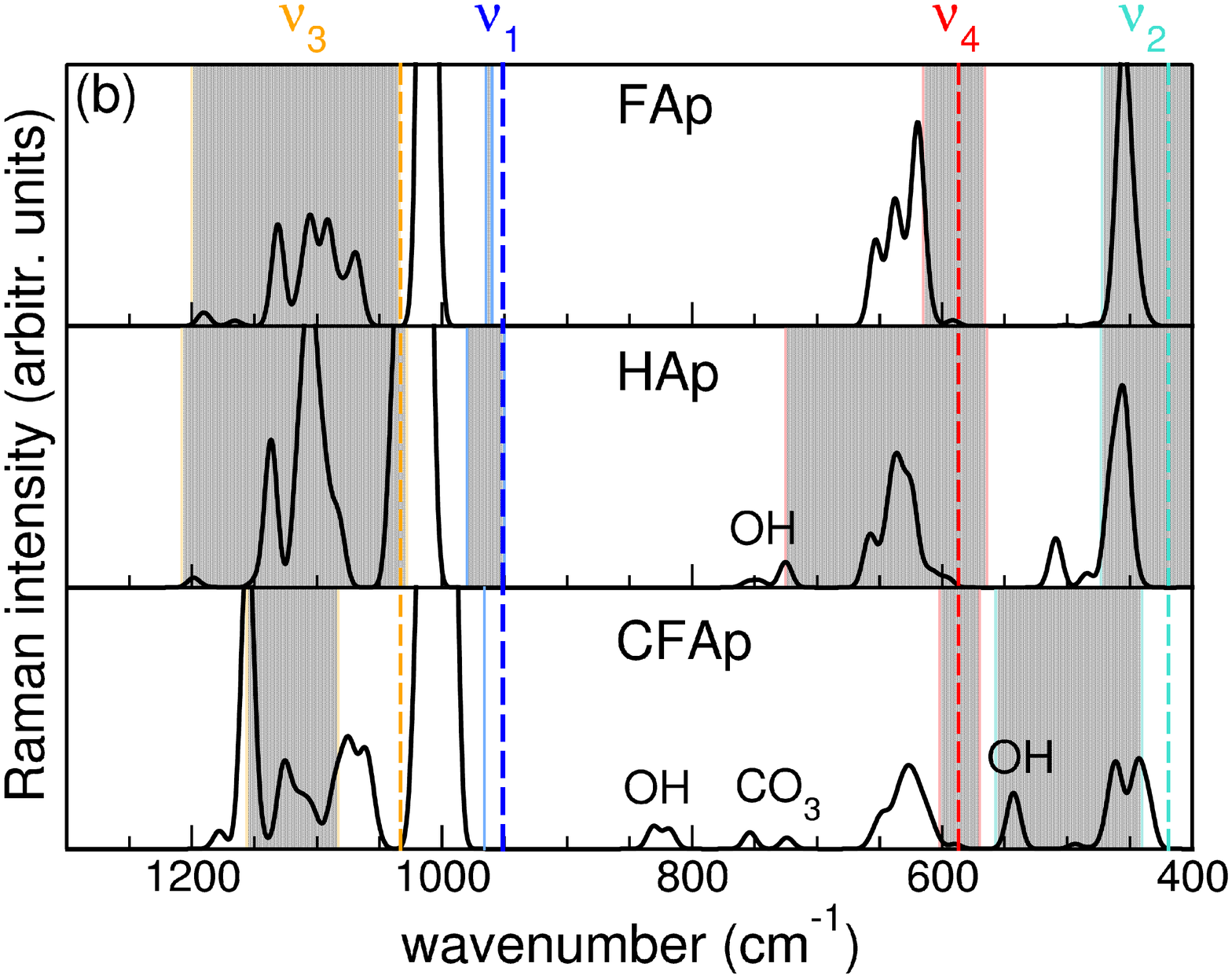}
 \caption{\small {\bf PO$_4^{3-}$ modes in apatite spectra.}
  (a) IR spectra and (b) Raman spectra of  FAp, HAp and CFAp in the frequency range $[400-1300]$ cm$^{-1}$.
  Shaded vertical bars  identify the total experimental range for the $\nu_1-\nu_4$ vibrational modes of orthophosphate units
  in apatites, taken from Table \ref{tab:var}. Dashed vertical lines mark the theoretical vibrational frequency of isolated PO$_4^{3-}$, as in Figure \ref{fig:2}b.
  Blue (turquoise) arrows point the $\nu_1$ ($\nu_2$) modes in apatites, not visible on this scale. All spectra include the
  LO-TO splitting contribution, the corresponding spectrum without LO-TO contribution (dark gray) is included in panel (a) for comparison. }
  \label{fig:3}
  \end{center}
\end{figure}

Starting from the vibrational and dielectric properties we simulated the IR and Raman spectra of the selected apatites
(Figures \ref{fig:3}). Polarized Raman spectra  along with the table with polarization selection rules for backscatteing geometries are reported in Figure S5 of SI.

Here, we focus on the frequency range $[400-1300]$ cm$^{-1}$, which corresponds to the phosphate bands;
lower energy modes $[0-400]$ cm$^{-1}$ will be discussed in the next Section.
The crystalline field  induces distortions in the PO$_4^{3-}$ tetrahedra that change the internal bond lengths and angles, and misaligns their orientation with
 respect to the ideal symmetry planes, giving  rise to shifts and splittings of the ideal PO$_4^{3-}$ normal modes.
 Consequently, the resulting internal PO$_4^{3-}$ vibrational bands depend on the actual crystallographic structure of each compound.
The latter is responsible for the huge variability in the definition of the phosphate vibrations and in the assignment of the single PO$_4^{3-}$ modes from the experiments. The presence of unintentional HPO$_4^{2-}$ impurities \cite{cusco98}, which implicitly involves a redistribution of the vibrational frequencies (see above), further complicates this analysis.

 Table \ref{tab:var} summarizes the present theoretical results  and a few experimental data from IR and Raman measurements.
 The comparison with the experiments makes evident a remarkable variation, not only for different apatites, but also within the same composition.
 The total range of variability for each PO$_4^{3-}$ mode and for each apatite is displayed for comparison as colored vertical bar
 in Figure \ref{fig:3}.

Despite such variations, the theoretical IR and Raman spectra are in general good agreement with the experimental measurements on
fluoro- \cite{leroy,klee,klee70} hydroxy- \cite{penel97,berry66,wei03} and carbonated-fluoro-apatites \cite{antonakos07},
well reproducing the energy position of the peaks and the their
relative intensity relations ({ i.e.} the shape of the spectra reproduces the experiments).
In particular, it is easy to distinguish for each apatite the same set  of four bands associated to the phosphate vibrations.
The theoretical results are slightly blue-shifted with respect to the experimental ones, especially for higher frequency bands $\nu_1$ and $\nu_3$.
This is a recurring behavior in {\it ab initio} phonon calculations, reported also in previous $\Gamma$-point simulation for HAp crystal \cite{corno06}.
In agreement with experiments \cite{leroy, wei03, antonakos07}, the  modes $\nu_1$ and $\nu_2$ are hardly detectable in IR spectra (see arrows in panel a).
This is reminiscent of the IR-inactive nature of the corresponding modes in the ideal PO$_4^{3-}$ anion. However, the structural disorder weakens this selection rule although
the IR intensity remains very low. With respect to the single molecule calculation (vertical dashed line in Figure \ref{fig:3}), both IR and Raman spectra show
the expected broadening and shifting of the PO$_4^{3-}$ modes, in agreement with the above discussion.
The analysis of the polarized Raman spectra (Figure S5, SI) allows also to assign the symmetry of the mode to Raman peaks. For example, it results
that for FAp, the $\nu_1$ peak has a A$_g$ character, while the $\nu_3$ band is a combination of two A$_g$ and three E$_g$ modes, in agreement with microRaman
polarized experiments \cite{leroy}.

\subsection*{Beyond the phosphate's modes}

The identification of the phosphate modes, albeit important in the apatite recognition, is not sufficient to explain the complexity of vibrational spectra of these minerals.
For instance, the long-range Coulomb interaction associated to the internal  polar character of apatites leads to a net LO-TO splitting to several phonon bands (Figure S4, SI).
 These contributions have an important  effect also on the vibrational spectra, as explicitly demonstrated in Figure \ref{fig:3}a for the FAp case (gray area),
 and are responsible, e.g., for the double-peaked shape of the  $\nu_3$ band, which is
 a distinctive characteristics of the IR spectra \cite{leroy}.
This feature cannot be justified on the basis of the PO$_4^{3-}$ vibrations, without taking into account the solid-state effects that result from the the overall components of the crystal.
Furthermore, since the PO$_4^{3-}$ anion are common to all apatite systems, their characterization is not useful to discriminate apatite with different composition.
As such, the phosphate vibrational modes cannot be exploited to identify the
different spectral features of HAp and CFAp  in the range $[500-800]$ cm$^{-1}$, which are associated to the hydroxy and carbonate units (Figure  \ref{fig:3}).

One alternative route to inspect the composition of apatite mineral is the analysis of the low frequency modes ($[0-400]$ cm$^{-1}$), often known as {\em lattice modes} since
they stem from the rotational modes of the PO$_4^{3-}$ groups along with the translational modes of the Ca$^{2+}$,  and PO$_4^{3-}$ sublattices and non-metal elements
(F$^-$, OH$^-$, CO$_3^-$, OH$^-$).
A part from the obvious three acoustic modes, all other modes in this range are generally complex and involve the displacement of all chemical species,
clearly seen in the atom-projected phonon density of states shown in Figure S4 of SI.
However, some characteristic phonons associated to single (not phosphate) elements  can be easily pinpointed.
On the experimental side, lattice modes were traditionally difficult to detect, and most papers published between '60s and '80s did not consider this frequency range.
Nowadays, the improvements in IR and Raman techniques allow for a direct investigation of these spectral features \cite{leroy,cusco98}.

\begin{figure}[!h!]
\begin{center}
\includegraphics[width=0.45\textwidth]{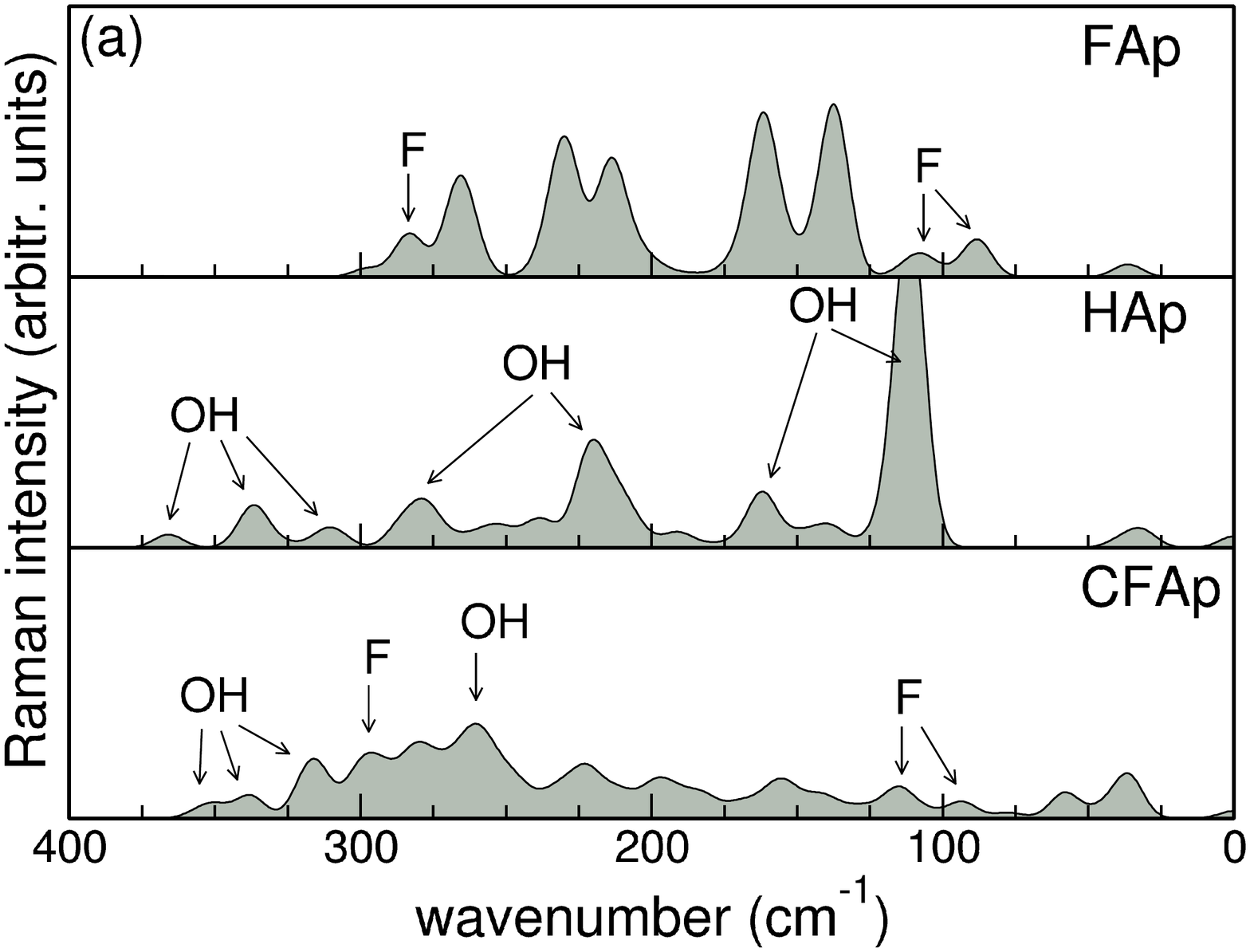}
\includegraphics[width=0.45\textwidth]{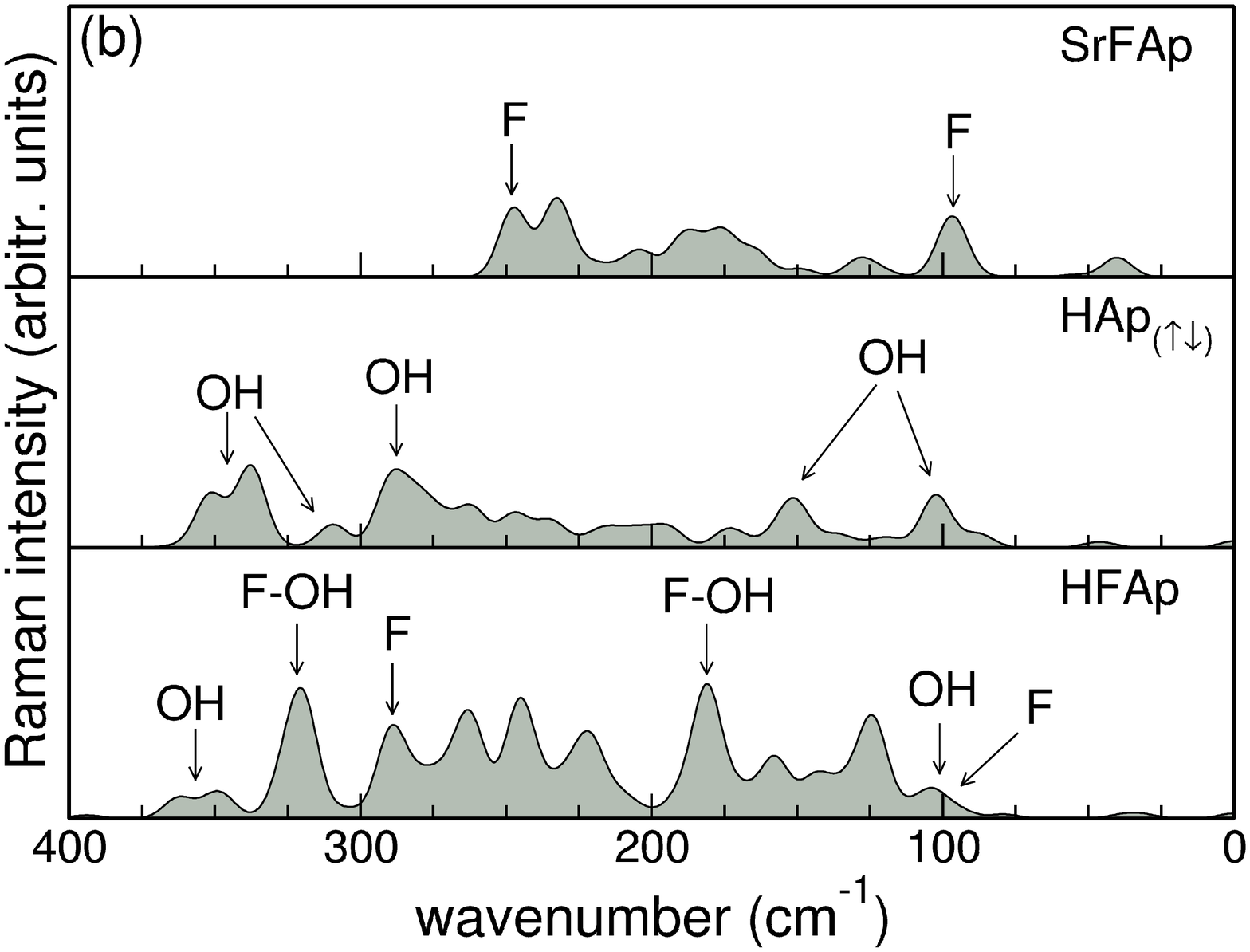}
  \caption{\small {\bf Lattice modes in apatite spectra.}
  Raman spectra in the frequency range $[0-400]$ cm$^{-1}$ of  (a) FAp, HAp and CFAp;  and (b)
 SrFAp, HAp$_{\uparrow\downarrow}$ and  HFAp hexagonal structures.}
  \label{fig:4}
  \end{center}
\end{figure}

From top to bottom panel, Figure \ref{fig:4}(a) shows the Raman spectra relative to the lattice modes of FAp, HAp, and CFAp, respectively.
First of all, we observe that, at odds with Figure \ref{fig:3} where all spectra had in common the same four-folded band structure,  the lattice modes have unique
spectral features depending on the composition. The lower is the symmetry of the crystal the broader is the spectrum.
Calcium, because of its larger mass, contributes to the majority of peaks in the range. However, since calcium is common to the three structures,
Ca-derived phonons are not useful to discriminate the three type of apatite. On the contrary, F and OH have specific features that can be isolated and used to fingerprint the chemical composition.
In fluoroapatite, F has characteristic peaks at 86, 107 and 282 cm$^{-1}$ (top panel). The same features are present at the same frequency also in CFAp, while are absent
in hydroxapatite. In a similar way, HAp and CFAp spectra have the same three-peaked structure in the range 309-364 cm$^{-1}$ associated to OH librational modes
in agreement with experimental results (335  cm$^{-1}$) \cite{cusco98}.
However, since in HAp and CFAp the number and the symmetry of OH ions are different, we also identify specific OH-derived peaks in the two spectra which account for
the different local environment.
Atomic contribution  are proved by the graphical inspection of the single phonon  modes and the simulation of Raman spectra with isotopic $^{16}$O/$^{18}$O and $^{19}$F/$^{18}$F
substitutions (Figure S6, SI). This proves that the Raman spectroscopy in the low-frequency region is a very powerful tool for the fine detection of the
apatite structure and composition.

\begin{figure}[!h!]
\begin{center}
\includegraphics[width=0.9\textwidth]{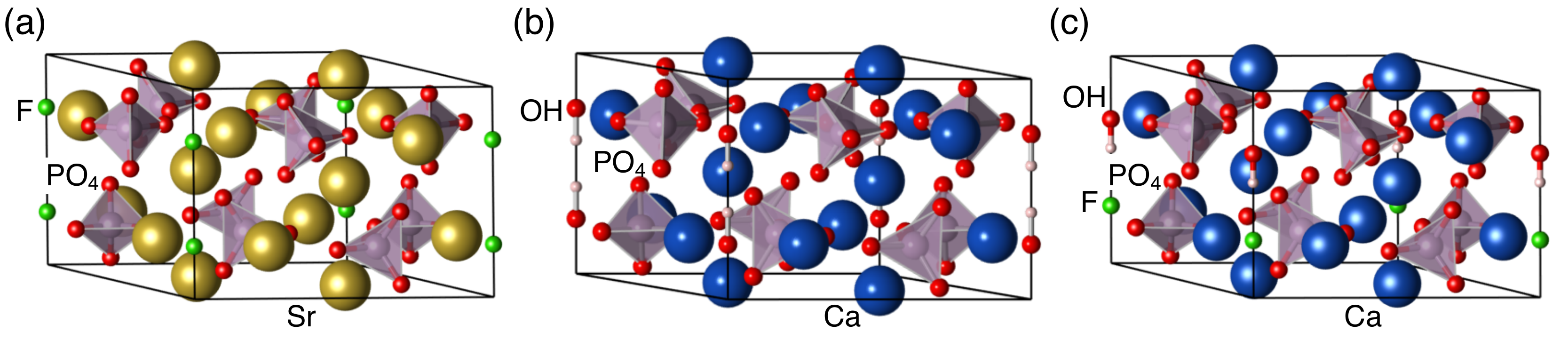}
  \caption{\small {\bf Crystal structure of alternative apatites.}
  3D view of hexagonal crystal cells of (a)  Sr-fluoroapatite (SrFAp), (b) antiparallel hydroxyapatite (HAp$_{\uparrow\downarrow}$) and (c)  hydroxy-fluoroapatite (HFAp).}
  \label{fig:5}
  \end{center}
\end{figure}

In order to further prove this statement, we considered  three alternative apatite crystals that have been experimentally studied,
namely Sr-fluoroapatite (SrFAp), antiparallel hydroxyapatite
(HAp$_{\uparrow\downarrow}$) and  hydroxy-fluoroapatite (HFAp), whose atomic structures are shown in Figure \ref{fig:5}.
SrFAp is obtained from FAp upon strontium substitution of Ca atoms.  The larger radius of Sr with respect to Ca causes an enlargement of the lattice parameters (a$_0$=10.05\AA, c=7.13\AA)  while the symmetry of the relaxed crystal ($S_6 (-3)$) is maintained.
HAp$_{(\uparrow\downarrow)}$ differs from HAp  for the spatial distribution of the OH ions that are
parallel $\uparrow\uparrow$ aligned (i.e. head-to-tail)
in HAp  and antiparallel  aligned $\uparrow\downarrow$ (i.e. tail-to-tail) in HAp$_{(\uparrow\downarrow})$.
The inversion of one OH ion increases the symmetry of the crystal (now $C_3 (3)$), due to the inversion
operation with respect to the glide plane. On the contrary, the substitution of one F with an OH ion in
HFAp breaks the internal symmetry of the crystal that reduces to C$_1$ point group.
By comparison with the three reference cases described above, these three examples helps in understanding: (i) the effect of Ca ion, (ii) the effect of order and alignment in
hydroxyapatites, and (iii) the effect of having mixed  F and OH element in the central Ca ion channels.
The polarized micro-Raman spectra are summarized in Figure S7 of SI.

The high frequency Raman spectra of SrFAp, HAp$_{(\uparrow\downarrow)}$, and HFAp are in agreement with the phosphate analysis discussed above,
the low energy spectra are shown in Figure \ref{fig:4}(b).
The larger mass of Sr with respect to Ca induces an overall shrinking and a red-shift of the spectrum (top panel). Two characteristic F peaks similar to FAp, can be easily
identified. The spatial inversion of one OH unit in hydroxyapatites (from  HAp$_{(\uparrow\uparrow})$ to  HAp$_{(\uparrow\downarrow})$) mostly affects the high-intensity peak of HAp at 109 cm$^{-1}$, associated to vertical OH-OH vibrational mode, which becomes strongly quenched in HAp$_{\uparrow\downarrow}$.
The identification  of this spectra features would allow to get information not only on the composition of the mineral (i.e. the presence of OH ions), but also 
on their local spatial distribution.
Finally the coexistence of both F and OH in the central Ca-channel shows the main fingerprints of the separate systems: F-derived peak at 96 and 287 cm$^{-1}$ and OH-derived
peaks at 102 and 302 cm$^{-1}$. Furthermore, new distinctive F-OH features now appear at 180 and 319 cm$^{-1}$, not observed in the other structures.

In summary, selected features in the IR and Raman spectra of calcium orthophosphates have been used extensively to characterize the chemical composition of biomininerals and analyze the interplay between chemistry, crystals nanoscale morphologies, and biological functions. The value of such spectroscopic techniques for {\it in vivo} and/or {\it in situ} analysis is unquestionable, especially in the research regarding calcified tissue deterioration. Other techniques, such as X-ray photoemission, may easily provide interpretation but are not generally applicable to monitor chemical changes, for instance, during fluorination.  We have shown that theoretical spectroscopies provide a solid tool to quantitatively fingerprint spectra and extract information until now mostly ignored. Our results point to the crucial contribution of the solvent in analyzing spectra of phosphate ions, show how phosphate modes are affected by surrounding ions, and present strategies to quantitatively detect the chemical composition of apatites by analyzing the full vibrational spectrum.
The synergy between first principles vibrational spectroscopies and experiments could facilitate the study of normal or pathological calcified tissues.

\section*{METHODS}

{\small Total-energy-and-forces optimization where carried out
within the density functional theory (DFT) framework, as  implemented in the Quantum
Espresso \cite{giannozzi17} (QE) suite of codes. The exchange-correlation functional is described by using the vdW-DF2-B86R \cite{Thonhauser15,Langreth15}
formulation of  generalized gradient approximation. The use of vdW-DF corrections to describe non-bonding  interactions was essential to stabilize the structure and lead to quantitatively agreement with available experimental data in apatites.
Ultrasoft pseudopotentials \cite {vanderbilt90} are used to treat the ionic potentials. Semicore {\em 3s3p} electrons of Ca are explicitly accounted in the valence shell.
Single particle Khon-Sham orbitals (charge densities) are expanded in plane waves up to a kinetic energy  cutoff of 40.0 Ry (800.0 Ry), respectively.
A uniform (4 $\times$ 4 $\times$ 6) k-point grid is used for summations over the Brillouin zone. The convergence thresholds for the geometrical
optimizations are set to $10^{-14}$ Ry for the total energy and $10^{-10}$ Ry/au for the forces.

Underestimation of the band gap with the standard DFT functionals is corrected by including a Hubbard-like potential on each chemical species, within the DFT+U
frameworks along the lines described, e.g., in Refs. \citenum{Anisimov00, Cococcioni05}. The optimized values for the studied compounds  are obtained by means of the
pseudohybrid Hubbard Density Functional approach (namely ACBN0) \cite{Agapito2015} and summarized in the SI (Tab. S1).
The effects of the ACBN0 approach on the structural, electronic and vibrational properties of inorganic semiconductors have been largely tested in previous reports
(see e.g. Refs. \citenum{calzolari13,Gopal:2015bf,eaton18}).

The phonon eigenvalues and eigenvectors are computed starting from the {\it  ab initio} calculation of the interatomic force matrices obtained
with a finite-differences/finite-fields approach \cite{calzolari13}, also implemented in the QE suite,
that has been demonstrated to be computationally very efficient in the simulation of the vibrational spectra of both inorganic \cite{calzolari13} and molecular crystals \cite{cigarini18}.
ACBN0, DFT and phonon calculations have been run by using the automatic workflows implemented in the AFLOW$\pi$  infrastructure \cite{aflowpi,paoflow}.

Molecular  anions  PO$_4^{3-}$ and HPO$_4^{2-}$  are simulated in cubic cells of dimension (14.0 $\times$ 14.0 $\times$ 14.0)\AA$^3$,
where a jellium background is inserted to remove divergences in the charged cells.
The dielectric effect of the environment (both water and apatite) on molecular phosphates is included through the {continuum solvation model} (CSM),\cite{tomasi}
implemented in the ENVIRON\cite{andreussi} add-on to QE.
The stability of the orthophosphate ion in water is simulated in an {explicit solvent model} (ESM), in which solvent molecules are treated on the same atomistic ground as the solute.
In this case, 3 Na$^+$ counter ions are included in the supercell to compensate the charges of phosphate, so that  the overall system (1 PO$_4^{3-}$ + 67 H$_2$O + 3 Na$^{+}$) is globally neutral.}

\section*{References}
\normalbaselines
\bibliography{biblio}

\begin{thebibliography}{69}%
\makeatletter
\providecommand \@ifxundefined [1]{%
 \@ifx{#1\undefined}
}%
\providecommand \@ifnum [1]{%
 \ifnum #1\expandafter \@firstoftwo
 \else \expandafter \@secondoftwo
 \fi
}%
\providecommand \@ifx [1]{%
 \ifx #1\expandafter \@firstoftwo
 \else \expandafter \@secondoftwo
 \fi
}%
\providecommand \natexlab [1]{#1}%
\providecommand \enquote  [1]{``#1''}%
\providecommand \bibnamefont  [1]{#1}%
\providecommand \bibfnamefont [1]{#1}%
\providecommand \citenamefont [1]{#1}%
\providecommand \href@noop [0]{\@secondoftwo}%
\providecommand \href [0]{\begingroup \@sanitize@url \@href}%
\providecommand \@href[1]{\@@startlink{#1}\@@href}%
\providecommand \@@href[1]{\endgroup#1\@@endlink}%
\providecommand \@sanitize@url [0]{\catcode `\\12\catcode `\$12\catcode
  `\&12\catcode `\#12\catcode `\^12\catcode `\_12\catcode `\%12\relax}%
\providecommand \@@startlink[1]{}%
\providecommand \@@endlink[0]{}%
\providecommand \url  [0]{\begingroup\@sanitize@url \@url }%
\providecommand \@url [1]{\endgroup\@href {#1}{\urlprefix }}%
\providecommand \urlprefix  [0]{URL }%
\providecommand \Eprint [0]{\href }%
\providecommand \doibase [0]{http://dx.doi.org/}%
\providecommand \selectlanguage [0]{\@gobble}%
\providecommand \bibinfo  [0]{\@secondoftwo}%
\providecommand \bibfield  [0]{\@secondoftwo}%
\providecommand \translation [1]{[#1]}%
\providecommand \BibitemOpen [0]{}%
\providecommand \bibitemStop [0]{}%
\providecommand \bibitemNoStop [0]{.\EOS\space}%
\providecommand \EOS [0]{\spacefactor3000\relax}%
\providecommand \BibitemShut  [1]{\csname bibitem#1\endcsname}%
\let\auto@bib@innerbib\@empty
\bibitem [{\citenamefont {Dorozhkin}(2009)}]{Dorozhkin2009}%
  \BibitemOpen
  \bibfield  {author} {\bibinfo {author} {\bibfnamefont {S.~V.}\ \bibnamefont
  {Dorozhkin}},\ }\href@noop {} {\bibfield  {journal} {\bibinfo  {journal}
  {Materials}\ }\textbf {\bibinfo {volume} {2}},\ \bibinfo {pages} {399}
  (\bibinfo {year} {2009})}\BibitemShut {NoStop}%
\bibitem [{\citenamefont {Iafisco}\ \emph {et~al.}(2013)\citenamefont
  {Iafisco}, \citenamefont {{Delgado-Lopez}}, \citenamefont {Varoni},
  \citenamefont {Tampieri}, \citenamefont {Rimondini}, \citenamefont
  {{Gomez-Morales}},\ and\ \citenamefont {Prat}}]{tampieri13}%
  \BibitemOpen
  \bibfield  {author} {\bibinfo {author} {\bibfnamefont {M.}~\bibnamefont
  {Iafisco}}, \bibinfo {author} {\bibfnamefont {J.~M.}\ \bibnamefont
  {{Delgado-Lopez}}}, \bibinfo {author} {\bibfnamefont {E.~M.}\ \bibnamefont
  {Varoni}}, \bibinfo {author} {\bibfnamefont {A.}~\bibnamefont {Tampieri}},
  \bibinfo {author} {\bibfnamefont {L.}~\bibnamefont {Rimondini}}, \bibinfo
  {author} {\bibfnamefont {J.}~\bibnamefont {{Gomez-Morales}}}, \ and\ \bibinfo
  {author} {\bibfnamefont {M.}~\bibnamefont {Prat}},\ }\href@noop {} {\bibfield
   {journal} {\bibinfo  {journal} {Small}\ }\textbf {\bibinfo {volume} {9}},\
  \bibinfo {pages} {3834} (\bibinfo {year} {2013})}\BibitemShut {NoStop}%
\bibitem [{\citenamefont {Faibish}\ \emph {et~al.}(2006)\citenamefont
  {Faibish}, \citenamefont {Ott},\ and\ \citenamefont {Boskey}}]{Faibish2006}%
  \BibitemOpen
  \bibfield  {author} {\bibinfo {author} {\bibfnamefont {D.}~\bibnamefont
  {Faibish}}, \bibinfo {author} {\bibfnamefont {S.~M.}\ \bibnamefont {Ott}}, \
  and\ \bibinfo {author} {\bibfnamefont {A.~L.}\ \bibnamefont {Boskey}},\
  }\href@noop {} {\bibfield  {journal} {\bibinfo  {journal} {Clin. Orthop.
  Relat. Res.}\ }\textbf {\bibinfo {volume} {443}},\ \bibinfo {pages} {28}
  (\bibinfo {year} {2006})}\BibitemShut {NoStop}%
\bibitem [{\citenamefont {Landi}\ \emph {et~al.}(2007)\citenamefont {Landi},
  \citenamefont {Tampieri}, \citenamefont {Celotti}, \citenamefont {Sprio},
  \citenamefont {Sandri},\ and\ \citenamefont {Logroscino}}]{tampieri07}%
  \BibitemOpen
  \bibfield  {author} {\bibinfo {author} {\bibfnamefont {E.}~\bibnamefont
  {Landi}}, \bibinfo {author} {\bibfnamefont {A.}~\bibnamefont {Tampieri}},
  \bibinfo {author} {\bibfnamefont {G.}~\bibnamefont {Celotti}}, \bibinfo
  {author} {\bibfnamefont {S.}~\bibnamefont {Sprio}}, \bibinfo {author}
  {\bibfnamefont {M.}~\bibnamefont {Sandri}}, \ and\ \bibinfo {author}
  {\bibfnamefont {G.}~\bibnamefont {Logroscino}},\ }\href@noop {} {\bibfield
  {journal} {\bibinfo  {journal} {Acta Biomaterialia}\ }\textbf {\bibinfo
  {volume} {3}},\ \bibinfo {pages} {961} (\bibinfo {year} {2007})}\BibitemShut
  {NoStop}%
\bibitem [{\citenamefont {{Al-Rekabi}}\ \emph {et~al.}(2017)\citenamefont
  {{Al-Rekabi}}, \citenamefont {Cunningham},\ and\ \citenamefont
  {Sniadecki}}]{AlRekabi2017}%
  \BibitemOpen
  \bibfield  {author} {\bibinfo {author} {\bibfnamefont {Z.}~\bibnamefont
  {{Al-Rekabi}}}, \bibinfo {author} {\bibfnamefont {M.~L.}\ \bibnamefont
  {Cunningham}}, \ and\ \bibinfo {author} {\bibfnamefont {N.~J.}\ \bibnamefont
  {Sniadecki}},\ }\href@noop {} {\bibfield  {journal} {\bibinfo  {journal} {ACS
  Biomat. Sci. Engin.}\ }\textbf {\bibinfo {volume} {3}},\ \bibinfo {pages}
  {2733} (\bibinfo {year} {2017})}\BibitemShut {NoStop}%
\bibitem [{\citenamefont {Landi}\ \emph {et~al.}(2000)\citenamefont {Landi},
  \citenamefont {Tampieri}, \citenamefont {Celotti},\ and\ \citenamefont
  {Sprio}}]{tampieri00}%
  \BibitemOpen
  \bibfield  {author} {\bibinfo {author} {\bibfnamefont {E.}~\bibnamefont
  {Landi}}, \bibinfo {author} {\bibfnamefont {A.}~\bibnamefont {Tampieri}},
  \bibinfo {author} {\bibfnamefont {G.}~\bibnamefont {Celotti}}, \ and\
  \bibinfo {author} {\bibfnamefont {S.}~\bibnamefont {Sprio}},\ }\href@noop {}
  {\bibfield  {journal} {\bibinfo  {journal} {J. Eur. Ceramic Soc.}\ }\textbf
  {\bibinfo {volume} {20}},\ \bibinfo {pages} {2377} (\bibinfo {year}
  {2000})}\BibitemShut {NoStop}%
\bibitem [{\citenamefont {{Rosin-Grget}}\ and\ \citenamefont
  {Lincir}(2001)}]{Rosin01}%
  \BibitemOpen
  \bibfield  {author} {\bibinfo {author} {\bibfnamefont {K.}~\bibnamefont
  {{Rosin-Grget}}}\ and\ \bibinfo {author} {\bibfnamefont {I.}~\bibnamefont
  {Lincir}},\ }\href@noop {} {\bibfield  {journal} {\bibinfo  {journal} {Coll
  Antropol.}\ }\textbf {\bibinfo {volume} {25}},\ \bibinfo {pages} {703}
  (\bibinfo {year} {2001})}\BibitemShut {NoStop}%
\bibitem [{\citenamefont {Arinzeh}\ \emph {et~al.}(2004)\citenamefont
  {Arinzeh}, \citenamefont {Tran}, \citenamefont {Mcalary},\ and\ \citenamefont
  {Daculsi}}]{Daculsi04}%
  \BibitemOpen
  \bibfield  {author} {\bibinfo {author} {\bibfnamefont {T.~L.}\ \bibnamefont
  {Arinzeh}}, \bibinfo {author} {\bibfnamefont {T.}~\bibnamefont {Tran}},
  \bibinfo {author} {\bibfnamefont {J.}~\bibnamefont {Mcalary}}, \ and\
  \bibinfo {author} {\bibfnamefont {G.}~\bibnamefont {Daculsi}},\ }\href@noop
  {} {\bibfield  {journal} {\bibinfo  {journal} {Biomaterials}\ }\textbf
  {\bibinfo {volume} {26}},\ \bibinfo {pages} {3631} (\bibinfo {year}
  {2004})}\BibitemShut {NoStop}%
\bibitem [{\citenamefont {LeGeros}(2008)}]{LeGeros08}%
  \BibitemOpen
  \bibfield  {author} {\bibinfo {author} {\bibfnamefont {R.~Z.}\ \bibnamefont
  {LeGeros}},\ }\href@noop {} {\bibfield  {journal} {\bibinfo  {journal} {Chem.
  Rev.}\ }\textbf {\bibinfo {volume} {108}},\ \bibinfo {pages} {4742} (\bibinfo
  {year} {2008})}\BibitemShut {NoStop}%
\bibitem [{\citenamefont {Tribble}\ \emph {et~al.}(1995)\citenamefont
  {Tribble}, \citenamefont {Arvidson}, \citenamefont {Lane},\ and\
  \citenamefont {Mackenzie}}]{Tribble95}%
  \BibitemOpen
  \bibfield  {author} {\bibinfo {author} {\bibfnamefont {J.~S.}\ \bibnamefont
  {Tribble}}, \bibinfo {author} {\bibfnamefont {R.~S.}\ \bibnamefont
  {Arvidson}}, \bibinfo {author} {\bibfnamefont {M.}~\bibnamefont {Lane}}, \
  and\ \bibinfo {author} {\bibfnamefont {F.~T.}\ \bibnamefont {Mackenzie}},\
  }\href@noop {} {\bibfield  {journal} {\bibinfo  {journal} {Sediment. Geol.}\
  }\textbf {\bibinfo {volume} {95}},\ \bibinfo {pages} {11} (\bibinfo {year}
  {1995})}\BibitemShut {NoStop}%
\bibitem [{\citenamefont {Sirbescu}\ \emph {et~al.}(2009)\citenamefont
  {Sirbescu}, \citenamefont {Leatherman}, \citenamefont {Student},\ and\
  \citenamefont {Beehr}}]{mona09}%
  \BibitemOpen
  \bibfield  {author} {\bibinfo {author} {\bibfnamefont {M.-L.~C.}\
  \bibnamefont {Sirbescu}}, \bibinfo {author} {\bibfnamefont {M.~A.}\
  \bibnamefont {Leatherman}}, \bibinfo {author} {\bibfnamefont {J.~J.}\
  \bibnamefont {Student}}, \ and\ \bibinfo {author} {\bibfnamefont {A.~R.}\
  \bibnamefont {Beehr}},\ }\href@noop {} {\bibfield  {journal} {\bibinfo
  {journal} {The Canadian Mineralogist}\ }\textbf {\bibinfo {volume} {47}},\
  \bibinfo {pages} {725} (\bibinfo {year} {2009})}\BibitemShut {NoStop}%
\bibitem [{\citenamefont {Zheng}\ \emph {et~al.}(2007)\citenamefont {Zheng},
  \citenamefont {Li}, \citenamefont {Yang}, \citenamefont {Zeng}, \citenamefont
  {Shen}, \citenamefont {Zhang},\ and\ \citenamefont {Liu}}]{zheng07}%
  \BibitemOpen
  \bibfield  {author} {\bibinfo {author} {\bibfnamefont {W.}~\bibnamefont
  {Zheng}}, \bibinfo {author} {\bibfnamefont {X.~M.}\ \bibnamefont {Li}},
  \bibinfo {author} {\bibfnamefont {Q.}~\bibnamefont {Yang}}, \bibinfo {author}
  {\bibfnamefont {G.~M.}\ \bibnamefont {Zeng}}, \bibinfo {author}
  {\bibfnamefont {X.~X.}\ \bibnamefont {Shen}}, \bibinfo {author}
  {\bibfnamefont {Y.}~\bibnamefont {Zhang}}, \ and\ \bibinfo {author}
  {\bibfnamefont {J.~J.}\ \bibnamefont {Liu}},\ }\href@noop {} {\bibfield
  {journal} {\bibinfo  {journal} {J. Hazard Mater.}\ }\textbf {\bibinfo
  {volume} {147}},\ \bibinfo {pages} {534} (\bibinfo {year}
  {2007})}\BibitemShut {NoStop}%
\bibitem [{\citenamefont {Wellman}\ \emph {et~al.}(2008)\citenamefont
  {Wellman}, \citenamefont {Glovack}, \citenamefont {Parker}, \citenamefont
  {Richards},\ and\ \citenamefont {Pierce}}]{wellman08}%
  \BibitemOpen
  \bibfield  {author} {\bibinfo {author} {\bibfnamefont {D.~M.}\ \bibnamefont
  {Wellman}}, \bibinfo {author} {\bibfnamefont {J.~N.}\ \bibnamefont
  {Glovack}}, \bibinfo {author} {\bibfnamefont {K.}~\bibnamefont {Parker}},
  \bibinfo {author} {\bibfnamefont {E.~L.}\ \bibnamefont {Richards}}, \ and\
  \bibinfo {author} {\bibfnamefont {E.~M.}\ \bibnamefont {Pierce}},\
  }\href@noop {} {\bibfield  {journal} {\bibinfo  {journal} {Environ. Chem.}\
  }\textbf {\bibinfo {volume} {5}},\ \bibinfo {pages} {40} (\bibinfo {year}
  {2008})}\BibitemShut {NoStop}%
\bibitem [{\citenamefont {Ensikat}\ \emph {et~al.}(2016)\citenamefont
  {Ensikat}, \citenamefont {Geisler},\ and\ \citenamefont
  {Weigend}}]{Weigend16}%
  \BibitemOpen
  \bibfield  {author} {\bibinfo {author} {\bibfnamefont {H.-J.}\ \bibnamefont
  {Ensikat}}, \bibinfo {author} {\bibfnamefont {T.}~\bibnamefont {Geisler}}, \
  and\ \bibinfo {author} {\bibfnamefont {M.}~\bibnamefont {Weigend}},\
  }\href@noop {} {\bibfield  {journal} {\bibinfo  {journal} {Sci. Rep.}\
  }\textbf {\bibinfo {volume} {6}},\ \bibinfo {pages} {26073} (\bibinfo {year}
  {2016})}\BibitemShut {NoStop}%
\bibitem [{\citenamefont {Weigend}\ \emph {et~al.}(2018)\citenamefont
  {Weigend}, \citenamefont {Mustafa},\ and\ \citenamefont
  {Ensikat}}]{Weigend18}%
  \BibitemOpen
  \bibfield  {author} {\bibinfo {author} {\bibfnamefont {M.}~\bibnamefont
  {Weigend}}, \bibinfo {author} {\bibfnamefont {A.}~\bibnamefont {Mustafa}}, \
  and\ \bibinfo {author} {\bibfnamefont {H.~J.}\ \bibnamefont {Ensikat}},\
  }\href@noop {} {\bibfield  {journal} {\bibinfo  {journal} {Planta}\ }\textbf
  {\bibinfo {volume} {247}},\ \bibinfo {pages} {277} (\bibinfo {year}
  {2018})}\BibitemShut {NoStop}%
\bibitem [{\citenamefont {Stifler}\ \emph {et~al.}(2018)\citenamefont
  {Stifler}, \citenamefont {Wittig}, \citenamefont {Sassi}, \citenamefont
  {Sun}, \citenamefont {Marcus}, \citenamefont {Birkedal}, \citenamefont
  {Beniash}, \citenamefont {Rosso},\ and\ \citenamefont {Gilbert}}]{stifler18}%
  \BibitemOpen
  \bibfield  {author} {\bibinfo {author} {\bibfnamefont {C.~A.}\ \bibnamefont
  {Stifler}}, \bibinfo {author} {\bibfnamefont {N.~K.}\ \bibnamefont {Wittig}},
  \bibinfo {author} {\bibfnamefont {M.}~\bibnamefont {Sassi}}, \bibinfo
  {author} {\bibfnamefont {C.-Y.}\ \bibnamefont {Sun}}, \bibinfo {author}
  {\bibfnamefont {M.~A.}\ \bibnamefont {Marcus}}, \bibinfo {author}
  {\bibfnamefont {H.}~\bibnamefont {Birkedal}}, \bibinfo {author}
  {\bibfnamefont {E.}~\bibnamefont {Beniash}}, \bibinfo {author} {\bibfnamefont
  {K.~M.}\ \bibnamefont {Rosso}}, \ and\ \bibinfo {author} {\bibfnamefont
  {P.~U. P.~A.}\ \bibnamefont {Gilbert}},\ }\href@noop {} {\bibfield  {journal}
  {\bibinfo  {journal} {J. Am. Chem. Soc.}\ }\textbf {\bibinfo {volume}
  {140}},\ \bibinfo {pages} {11698} (\bibinfo {year} {2018})}\BibitemShut
  {NoStop}%
\bibitem [{\citenamefont {Marcus}\ \emph {et~al.}(2017)\citenamefont {Marcus},
  \citenamefont {Amini}, \citenamefont {Stifler}, \citenamefont {Sun},
  \citenamefont {Tamura}, \citenamefont {Bechtel}, \citenamefont {Parkinson},
  \citenamefont {Barnard}, \citenamefont {Zhang}, \citenamefont {Chua},
  \citenamefont {Miserez},\ and\ \citenamefont {Gilbert}}]{stifler17}%
  \BibitemOpen
  \bibfield  {author} {\bibinfo {author} {\bibfnamefont {M.~A.}\ \bibnamefont
  {Marcus}}, \bibinfo {author} {\bibfnamefont {S.}~\bibnamefont {Amini}},
  \bibinfo {author} {\bibfnamefont {C.~A.}\ \bibnamefont {Stifler}}, \bibinfo
  {author} {\bibfnamefont {C.-Y.}\ \bibnamefont {Sun}}, \bibinfo {author}
  {\bibfnamefont {N.}~\bibnamefont {Tamura}}, \bibinfo {author} {\bibfnamefont
  {H.~A.}\ \bibnamefont {Bechtel}}, \bibinfo {author} {\bibfnamefont {D.~Y.}\
  \bibnamefont {Parkinson}}, \bibinfo {author} {\bibfnamefont {H.~S.}\
  \bibnamefont {Barnard}}, \bibinfo {author} {\bibfnamefont {X.~X.~X.}\
  \bibnamefont {Zhang}}, \bibinfo {author} {\bibfnamefont {J.~Q.~I.}\
  \bibnamefont {Chua}}, \bibinfo {author} {\bibfnamefont {A.}~\bibnamefont
  {Miserez}}, \ and\ \bibinfo {author} {\bibfnamefont {P.~U. P.~A.}\
  \bibnamefont {Gilbert}},\ }\href@noop {} {\bibfield  {journal} {\bibinfo
  {journal} {ACS Nano}\ }\textbf {\bibinfo {volume} {11}},\ \bibinfo {pages}
  {11856} (\bibinfo {year} {2017})}\BibitemShut {NoStop}%
\bibitem [{\citenamefont {Lotsari}\ \emph {et~al.}(2018)\citenamefont
  {Lotsari}, \citenamefont {Rajasekharan}, \citenamefont {Halvarsson},\ and\
  \citenamefont {Andersson}}]{lotsari18}%
  \BibitemOpen
  \bibfield  {author} {\bibinfo {author} {\bibfnamefont {A.}~\bibnamefont
  {Lotsari}}, \bibinfo {author} {\bibfnamefont {A.~K.}\ \bibnamefont
  {Rajasekharan}}, \bibinfo {author} {\bibfnamefont {M.}~\bibnamefont
  {Halvarsson}}, \ and\ \bibinfo {author} {\bibfnamefont {M.}~\bibnamefont
  {Andersson}},\ }\href@noop {} {\bibfield  {journal} {\bibinfo  {journal}
  {Nature Commun.}\ }\textbf {\bibinfo {volume} {9}},\ \bibinfo {pages} {4170}
  (\bibinfo {year} {2018})}\BibitemShut {NoStop}%
\bibitem [{\citenamefont {Nudelman}\ \emph {et~al.}(2010)\citenamefont
  {Nudelman}, \citenamefont {Pieterse}, \citenamefont {George}, \citenamefont
  {Bomans}, \citenamefont {Friedrich}, \citenamefont {Brylka}, \citenamefont
  {Hilbers}, \citenamefont {{de With}},\ and\ \citenamefont
  {Sommerdijk}}]{Nudelman10}%
  \BibitemOpen
  \bibfield  {author} {\bibinfo {author} {\bibfnamefont {F.}~\bibnamefont
  {Nudelman}}, \bibinfo {author} {\bibfnamefont {K.}~\bibnamefont {Pieterse}},
  \bibinfo {author} {\bibfnamefont {A.}~\bibnamefont {George}}, \bibinfo
  {author} {\bibfnamefont {P.~H.~H.}\ \bibnamefont {Bomans}}, \bibinfo {author}
  {\bibfnamefont {H.}~\bibnamefont {Friedrich}}, \bibinfo {author}
  {\bibfnamefont {L.~J.}\ \bibnamefont {Brylka}}, \bibinfo {author}
  {\bibfnamefont {P.~A.~J.}\ \bibnamefont {Hilbers}}, \bibinfo {author}
  {\bibfnamefont {G.}~\bibnamefont {{de With}}}, \ and\ \bibinfo {author}
  {\bibfnamefont {N.~A. J.~M.}\ \bibnamefont {Sommerdijk}},\ }\href@noop {}
  {\bibfield  {journal} {\bibinfo  {journal} {Nature Mater.}\ }\textbf
  {\bibinfo {volume} {9}},\ \bibinfo {pages} {1004} (\bibinfo {year}
  {2010})}\BibitemShut {NoStop}%
\bibitem [{\citenamefont {Suchaneka}\ \emph {et~al.}(2004)\citenamefont
  {Suchaneka}, \citenamefont {Byrappa}, \citenamefont {Shuk}, \citenamefont
  {Riman}, \citenamefont {Janas},\ and\ \citenamefont
  {TenHuisen}}]{Suchanek04}%
  \BibitemOpen
  \bibfield  {author} {\bibinfo {author} {\bibfnamefont {W.~L.}\ \bibnamefont
  {Suchaneka}}, \bibinfo {author} {\bibfnamefont {K.}~\bibnamefont {Byrappa}},
  \bibinfo {author} {\bibfnamefont {P.}~\bibnamefont {Shuk}}, \bibinfo {author}
  {\bibfnamefont {R.~E.}\ \bibnamefont {Riman}}, \bibinfo {author}
  {\bibfnamefont {V.~F.}\ \bibnamefont {Janas}}, \ and\ \bibinfo {author}
  {\bibfnamefont {K.~S.}\ \bibnamefont {TenHuisen}},\ }\href@noop {} {\bibfield
   {journal} {\bibinfo  {journal} {Biomaterials}\ }\textbf {\bibinfo {volume}
  {25}},\ \bibinfo {pages} {4647} (\bibinfo {year} {2004})}\BibitemShut
  {NoStop}%
\bibitem [{\citenamefont {Elkabouss}\ \emph {et~al.}(2004)\citenamefont
  {Elkabouss}, \citenamefont {Kacimi}, \citenamefont {Ziyad}, \citenamefont
  {Ammar},\ and\ \citenamefont {{Bozon-Verduraz}}}]{Elkabouss04}%
  \BibitemOpen
  \bibfield  {author} {\bibinfo {author} {\bibfnamefont {K.}~\bibnamefont
  {Elkabouss}}, \bibinfo {author} {\bibfnamefont {M.}~\bibnamefont {Kacimi}},
  \bibinfo {author} {\bibfnamefont {M.}~\bibnamefont {Ziyad}}, \bibinfo
  {author} {\bibfnamefont {S.}~\bibnamefont {Ammar}}, \ and\ \bibinfo {author}
  {\bibfnamefont {F.}~\bibnamefont {{Bozon-Verduraz}}},\ }\href@noop {}
  {\bibfield  {journal} {\bibinfo  {journal} {J. Catalysis}\ }\textbf {\bibinfo
  {volume} {226}},\ \bibinfo {pages} {16} (\bibinfo {year} {2004})}\BibitemShut
  {NoStop}%
\bibitem [{\citenamefont {Ohtsu}\ \emph {et~al.}(2013)\citenamefont {Ohtsu},
  \citenamefont {Hiromoto}, \citenamefont {Yamane}, \citenamefont {Satoh},\
  and\ \citenamefont {Tomozawa}}]{Ohtsu13}%
  \BibitemOpen
  \bibfield  {author} {\bibinfo {author} {\bibfnamefont {N.}~\bibnamefont
  {Ohtsu}}, \bibinfo {author} {\bibfnamefont {S.}~\bibnamefont {Hiromoto}},
  \bibinfo {author} {\bibfnamefont {M.}~\bibnamefont {Yamane}}, \bibinfo
  {author} {\bibfnamefont {K.}~\bibnamefont {Satoh}}, \ and\ \bibinfo {author}
  {\bibfnamefont {M.}~\bibnamefont {Tomozawa}},\ }\href@noop {} {\bibfield
  {journal} {\bibinfo  {journal} {Surf. Coat. Technol.}\ }\textbf {\bibinfo
  {volume} {218}},\ \bibinfo {pages} {114} (\bibinfo {year}
  {2013})}\BibitemShut {NoStop}%
\bibitem [{\citenamefont {Kruse}\ \emph {et~al.}(2009)\citenamefont {Kruse},
  \citenamefont {Leinweber}, \citenamefont {Eckhardt}, \citenamefont
  {Godlinski}, \citenamefont {Huc},\ and\ \citenamefont {Zuin}}]{kruse09}%
  \BibitemOpen
  \bibfield  {author} {\bibinfo {author} {\bibfnamefont {J.}~\bibnamefont
  {Kruse}}, \bibinfo {author} {\bibfnamefont {P.}~\bibnamefont {Leinweber}},
  \bibinfo {author} {\bibfnamefont {K.-U.}\ \bibnamefont {Eckhardt}}, \bibinfo
  {author} {\bibfnamefont {F.}~\bibnamefont {Godlinski}}, \bibinfo {author}
  {\bibfnamefont {Y.}~\bibnamefont {Huc}}, \ and\ \bibinfo {author}
  {\bibfnamefont {L.}~\bibnamefont {Zuin}},\ }\href@noop {} {\bibfield
  {journal} {\bibinfo  {journal} {J. Synchrotron Rad.}\ }\textbf {\bibinfo
  {volume} {16}},\ \bibinfo {pages} {247} (\bibinfo {year} {2009})}\BibitemShut
  {NoStop}%
\bibitem [{\citenamefont {Nakata}\ \emph {et~al.}(2009)\citenamefont {Nakata},
  \citenamefont {Kubo}, \citenamefont {Numako}, \citenamefont {Onoki},\ and\
  \citenamefont {Nakahir}}]{nakata09}%
  \BibitemOpen
  \bibfield  {author} {\bibinfo {author} {\bibfnamefont {K.}~\bibnamefont
  {Nakata}}, \bibinfo {author} {\bibfnamefont {T.}~\bibnamefont {Kubo}},
  \bibinfo {author} {\bibfnamefont {C.}~\bibnamefont {Numako}}, \bibinfo
  {author} {\bibfnamefont {T.}~\bibnamefont {Onoki}}, \ and\ \bibinfo {author}
  {\bibfnamefont {A.}~\bibnamefont {Nakahir}},\ }\href@noop {} {\bibfield
  {journal} {\bibinfo  {journal} {Mater. Trans.}\ }\textbf {\bibinfo {volume}
  {50}},\ \bibinfo {pages} {1046} (\bibinfo {year} {2009})}\BibitemShut
  {NoStop}%
\bibitem [{\citenamefont {Kashiwabara}\ \emph {et~al.}(2014)\citenamefont
  {Kashiwabara}, \citenamefont {Toda}, \citenamefont {Fujinaga}, \citenamefont
  {Honma}, \citenamefont {Takahashi},\ and\ \citenamefont {Kato}}]{kato14}%
  \BibitemOpen
  \bibfield  {author} {\bibinfo {author} {\bibfnamefont {T.}~\bibnamefont
  {Kashiwabara}}, \bibinfo {author} {\bibfnamefont {R.}~\bibnamefont {Toda}},
  \bibinfo {author} {\bibfnamefont {K.}~\bibnamefont {Fujinaga}}, \bibinfo
  {author} {\bibfnamefont {T.}~\bibnamefont {Honma}}, \bibinfo {author}
  {\bibfnamefont {Y.}~\bibnamefont {Takahashi}}, \ and\ \bibinfo {author}
  {\bibfnamefont {Y.}~\bibnamefont {Kato}},\ }\href@noop {} {\bibfield
  {journal} {\bibinfo  {journal} {Chem. Lett.}\ }\textbf {\bibinfo {volume}
  {43}},\ \bibinfo {pages} {199} (\bibinfo {year} {2014})}\BibitemShut
  {NoStop}%
\bibitem [{\citenamefont {Pan}\ and\ \citenamefont {Fleet}(2002)}]{pan02}%
  \BibitemOpen
  \bibfield  {author} {\bibinfo {author} {\bibfnamefont {Y.~M.}\ \bibnamefont
  {Pan}}\ and\ \bibinfo {author} {\bibfnamefont {M.~E.}\ \bibnamefont
  {Fleet}},\ }\href@noop {} {\bibfield  {journal} {\bibinfo  {journal}
  {Phosphates: Geochemical, Geobiological, And Materials Importance Book
  Series: Reviews In Mineralogy And Geochemistry}\ }\textbf {\bibinfo {volume}
  {48}},\ \bibinfo {pages} {13} (\bibinfo {year} {2002})}\BibitemShut {NoStop}%
\bibitem [{\citenamefont {Panda}\ \emph {et~al.}(2003)\citenamefont {Panda},
  \citenamefont {Hsieh}, \citenamefont {Chung},\ and\ \citenamefont
  {Chin}}]{panda03}%
  \BibitemOpen
  \bibfield  {author} {\bibinfo {author} {\bibfnamefont {R.~N.}\ \bibnamefont
  {Panda}}, \bibinfo {author} {\bibfnamefont {M.~F.}\ \bibnamefont {Hsieh}},
  \bibinfo {author} {\bibfnamefont {R.~J.}\ \bibnamefont {Chung}}, \ and\
  \bibinfo {author} {\bibfnamefont {T.}~\bibnamefont {Chin}},\ }\href@noop {}
  {\bibfield  {journal} {\bibinfo  {journal} {J. Phys. Chem. Sol.}\ }\textbf
  {\bibinfo {volume} {64}},\ \bibinfo {pages} {193} (\bibinfo {year}
  {2003})}\BibitemShut {NoStop}%
\bibitem [{\citenamefont {Hu}\ \emph {et~al.}(2010)\citenamefont {Hu},
  \citenamefont {Rawal},\ and\ \citenamefont {{Schmidt-Rohr}}}]{hu10}%
  \BibitemOpen
  \bibfield  {author} {\bibinfo {author} {\bibfnamefont {Y.-Y.}\ \bibnamefont
  {Hu}}, \bibinfo {author} {\bibfnamefont {A.}~\bibnamefont {Rawal}}, \ and\
  \bibinfo {author} {\bibfnamefont {K.}~\bibnamefont {{Schmidt-Rohr}}},\
  }\href@noop {} {\bibfield  {journal} {\bibinfo  {journal} {Proc. Natl. Acc.
  Sci.}\ }\textbf {\bibinfo {volume} {107}},\ \bibinfo {pages} {22425}
  (\bibinfo {year} {2010})}\BibitemShut {NoStop}%
\bibitem [{\citenamefont {Pavan}\ \emph {et~al.}(2012)\citenamefont {Pavan},
  \citenamefont {Ceresoli}, \citenamefont {Tecklenburg},\ and\ \citenamefont
  {Fornari}}]{pavan12}%
  \BibitemOpen
  \bibfield  {author} {\bibinfo {author} {\bibfnamefont {B.}~\bibnamefont
  {Pavan}}, \bibinfo {author} {\bibfnamefont {D.}~\bibnamefont {Ceresoli}},
  \bibinfo {author} {\bibfnamefont {M.~M.~J.}\ \bibnamefont {Tecklenburg}}, \
  and\ \bibinfo {author} {\bibfnamefont {M.}~\bibnamefont {Fornari}},\
  }\href@noop {} {\bibfield  {journal} {\bibinfo  {journal} {Sol. St. Nuclear
  Mag. Res.}\ }\textbf {\bibinfo {volume} {45--46}},\ \bibinfo {pages} {59}
  (\bibinfo {year} {2012})}\BibitemShut {NoStop}%
\bibitem [{\citenamefont {Schulze}\ \emph {et~al.}(1973)\citenamefont
  {Schulze}, \citenamefont {Weinstock}, \citenamefont {M\"uller},\ and\
  \citenamefont {Vandrish}}]{schultze73}%
  \BibitemOpen
  \bibfield  {author} {\bibinfo {author} {\bibfnamefont {H.}~\bibnamefont
  {Schulze}}, \bibinfo {author} {\bibfnamefont {N.}~\bibnamefont {Weinstock}},
  \bibinfo {author} {\bibfnamefont {A.}~\bibnamefont {M\"uller}}, \ and\
  \bibinfo {author} {\bibfnamefont {G.}~\bibnamefont {Vandrish}},\ }\href@noop
  {} {\bibfield  {journal} {\bibinfo  {journal} {Spectrochim. Acta}\ }\textbf
  {\bibinfo {volume} {29A}},\ \bibinfo {pages} {1705} (\bibinfo {year}
  {1973})}\BibitemShut {NoStop}%
\bibitem [{\citenamefont {Mason}\ \emph {et~al.}(2003)\citenamefont {Mason},
  \citenamefont {Cruickshank}, \citenamefont {Neilson},\ and\ \citenamefont
  {Buchanan}}]{mason03}%
  \BibitemOpen
  \bibfield  {author} {\bibinfo {author} {\bibfnamefont {P.~E.}\ \bibnamefont
  {Mason}}, \bibinfo {author} {\bibfnamefont {J.~M.}\ \bibnamefont
  {Cruickshank}}, \bibinfo {author} {\bibfnamefont {G.~W.}\ \bibnamefont
  {Neilson}}, \ and\ \bibinfo {author} {\bibfnamefont {P.}~\bibnamefont
  {Buchanan}},\ }\href@noop {} {\bibfield  {journal} {\bibinfo  {journal}
  {Phys. Chem. Chem. Phys.}\ }\textbf {\bibinfo {volume} {5}},\ \bibinfo
  {pages} {4686} (\bibinfo {year} {2003})}\BibitemShut {NoStop}%
\bibitem [{\citenamefont {Chapman}\ and\ \citenamefont
  {Thirlwell}(1964)}]{chapman64}%
  \BibitemOpen
  \bibfield  {author} {\bibinfo {author} {\bibfnamefont {A.~C.}\ \bibnamefont
  {Chapman}}\ and\ \bibinfo {author} {\bibfnamefont {L.~E.}\ \bibnamefont
  {Thirlwell}},\ }\href@noop {} {\bibfield  {journal} {\bibinfo  {journal}
  {Spectrochim. Acta}\ }\textbf {\bibinfo {volume} {20}},\ \bibinfo {pages}
  {937} (\bibinfo {year} {1964})}\BibitemShut {NoStop}%
\bibitem [{\citenamefont {Preston}\ and\ \citenamefont
  {Adams}(1979)}]{preston79}%
  \BibitemOpen
  \bibfield  {author} {\bibinfo {author} {\bibfnamefont {C.~M.}\ \bibnamefont
  {Preston}}\ and\ \bibinfo {author} {\bibfnamefont {W.~A.}\ \bibnamefont
  {Adams}},\ }\href@noop {} {\bibfield  {journal} {\bibinfo  {journal} {J.
  Chem. Phys.}\ }\textbf {\bibinfo {volume} {83}},\ \bibinfo {pages} {1814}
  (\bibinfo {year} {1979})}\BibitemShut {NoStop}%
\bibitem [{\citenamefont {Cherif}\ \emph {et~al.}(2000)\citenamefont {Cherif},
  \citenamefont {Mgaidi}, \citenamefont {Ammar}, \citenamefont {Vallee},\ and\
  \citenamefont {F\"urst}}]{cherif00}%
  \BibitemOpen
  \bibfield  {author} {\bibinfo {author} {\bibfnamefont {M.}~\bibnamefont
  {Cherif}}, \bibinfo {author} {\bibfnamefont {A.}~\bibnamefont {Mgaidi}},
  \bibinfo {author} {\bibfnamefont {N.}~\bibnamefont {Ammar}}, \bibinfo
  {author} {\bibfnamefont {G.}~\bibnamefont {Vallee}}, \ and\ \bibinfo {author}
  {\bibfnamefont {W.}~\bibnamefont {F\"urst}},\ }\href@noop {} {\bibfield
  {journal} {\bibinfo  {journal} {J. Sol. Chem.}\ }\textbf {\bibinfo {volume}
  {29}},\ \bibinfo {pages} {255} (\bibinfo {year} {2000})}\BibitemShut
  {NoStop}%
\bibitem [{\citenamefont {Rudolph}(2010)}]{rudolph10}%
  \BibitemOpen
  \bibfield  {author} {\bibinfo {author} {\bibfnamefont {W.~W.}\ \bibnamefont
  {Rudolph}},\ }\href@noop {} {\bibfield  {journal} {\bibinfo  {journal}
  {Dalton Trans.}\ }\textbf {\bibinfo {volume} {39}},\ \bibinfo {pages} {9642}
  (\bibinfo {year} {2010})}\BibitemShut {NoStop}%
\bibitem [{\citenamefont {Syed}\ \emph {et~al.}(2012)\citenamefont {Syed},
  \citenamefont {Pang}, \citenamefont {Zhang},\ and\ \citenamefont
  {Zhang}}]{ajmal12}%
  \BibitemOpen
  \bibfield  {author} {\bibinfo {author} {\bibfnamefont {K.~A.}\ \bibnamefont
  {Syed}}, \bibinfo {author} {\bibfnamefont {S.-F.}\ \bibnamefont {Pang}},
  \bibinfo {author} {\bibfnamefont {Y.}~\bibnamefont {Zhang}}, \ and\ \bibinfo
  {author} {\bibfnamefont {Y.-H.}\ \bibnamefont {Zhang}},\ }\href@noop {}
  {\bibfield  {journal} {\bibinfo  {journal} {J. Phys. Chem. A}\ }\textbf
  {\bibinfo {volume} {116}},\ \bibinfo {pages} {1558} (\bibinfo {year}
  {2012})}\BibitemShut {NoStop}%
\bibitem [{\citenamefont {Syed}\ \emph {et~al.}(2013)\citenamefont {Syed},
  \citenamefont {Pang}, \citenamefont {Zhang},\ and\ \citenamefont
  {Zhang}}]{ajmal13}%
  \BibitemOpen
  \bibfield  {author} {\bibinfo {author} {\bibfnamefont {K.~A.}\ \bibnamefont
  {Syed}}, \bibinfo {author} {\bibfnamefont {S.-F.}\ \bibnamefont {Pang}},
  \bibinfo {author} {\bibfnamefont {Y.}~\bibnamefont {Zhang}}, \ and\ \bibinfo
  {author} {\bibfnamefont {Y.-H.}\ \bibnamefont {Zhang}},\ }\href@noop {}
  {\bibfield  {journal} {\bibinfo  {journal} {J. Chem. Phys.}\ }\textbf
  {\bibinfo {volume} {138}},\ \bibinfo {pages} {024901} (\bibinfo {year}
  {2013})}\BibitemShut {NoStop}%
\bibitem [{\citenamefont {Kl\"ahn}\ \emph {et~al.}(2004)\citenamefont
  {Kl\"ahn}, \citenamefont {Mathias}, \citenamefont {K\:otting}, \citenamefont
  {Nonella}, \citenamefont {Schlitter}, \citenamefont {Gerwert},\ and\
  \citenamefont {Tavan}}]{tavan04}%
  \BibitemOpen
  \bibfield  {author} {\bibinfo {author} {\bibfnamefont {M.}~\bibnamefont
  {Kl\"ahn}}, \bibinfo {author} {\bibfnamefont {G.}~\bibnamefont {Mathias}},
  \bibinfo {author} {\bibfnamefont {C.}~\bibnamefont {K\:otting}}, \bibinfo
  {author} {\bibfnamefont {M.}~\bibnamefont {Nonella}}, \bibinfo {author}
  {\bibfnamefont {J.}~\bibnamefont {Schlitter}}, \bibinfo {author}
  {\bibfnamefont {K.}~\bibnamefont {Gerwert}}, \ and\ \bibinfo {author}
  {\bibfnamefont {P.}~\bibnamefont {Tavan}},\ }\href@noop {} {\bibfield
  {journal} {\bibinfo  {journal} {J. Phys. Chem. A}\ }\textbf {\bibinfo
  {volume} {108}},\ \bibinfo {pages} {6186} (\bibinfo {year}
  {2004})}\BibitemShut {NoStop}%
\bibitem [{\citenamefont {Ebner}\ \emph {et~al.}(2005)\citenamefont {Ebner},
  \citenamefont {Onthong},\ and\ \citenamefont {Probst}}]{ebner05}%
  \BibitemOpen
  \bibfield  {author} {\bibinfo {author} {\bibfnamefont {C.}~\bibnamefont
  {Ebner}}, \bibinfo {author} {\bibfnamefont {U.}~\bibnamefont {Onthong}}, \
  and\ \bibinfo {author} {\bibfnamefont {M.}~\bibnamefont {Probst}},\
  }\href@noop {} {\bibfield  {journal} {\bibinfo  {journal} {J. Mol. Liq.}\
  }\textbf {\bibinfo {volume} {118}},\ \bibinfo {pages} {15} (\bibinfo {year}
  {2005})}\BibitemShut {NoStop}%
\bibitem [{\citenamefont {Corno}\ \emph {et~al.}(2006)\citenamefont {Corno},
  \citenamefont {Busco}, \citenamefont {Civalleria},\ and\ \citenamefont
  {Ugliengo}}]{corno06}%
  \BibitemOpen
  \bibfield  {author} {\bibinfo {author} {\bibfnamefont {M.}~\bibnamefont
  {Corno}}, \bibinfo {author} {\bibfnamefont {C.}~\bibnamefont {Busco}},
  \bibinfo {author} {\bibfnamefont {B.}~\bibnamefont {Civalleria}}, \ and\
  \bibinfo {author} {\bibfnamefont {P.}~\bibnamefont {Ugliengo}},\ }\href@noop
  {} {\bibfield  {journal} {\bibinfo  {journal} {Phys. Chem. Chem. Phys.}\
  }\textbf {\bibinfo {volume} {8}},\ \bibinfo {pages} {2464} (\bibinfo {year}
  {2006})}\BibitemShut {NoStop}%
\bibitem [{\citenamefont {Saenger}\ and\ \citenamefont
  {Kuhs}(1992)}]{saenger92}%
  \BibitemOpen
  \bibfield  {author} {\bibinfo {author} {\bibfnamefont {A.~T.}\ \bibnamefont
  {Saenger}}\ and\ \bibinfo {author} {\bibfnamefont {W.~F.}\ \bibnamefont
  {Kuhs}},\ }\href@noop {} {\bibfield  {journal} {\bibinfo  {journal} {Z.
  Kristallogr.}\ }\textbf {\bibinfo {volume} {199}},\ \bibinfo {pages} {123}
  (\bibinfo {year} {1992})}\BibitemShut {NoStop}%
\bibitem [{\citenamefont {Fleet}\ and\ \citenamefont {Pan}(1995)}]{fleet95}%
  \BibitemOpen
  \bibfield  {author} {\bibinfo {author} {\bibfnamefont {M.~E.}\ \bibnamefont
  {Fleet}}\ and\ \bibinfo {author} {\bibfnamefont {Y.}~\bibnamefont {Pan}},\
  }\href@noop {} {\bibfield  {journal} {\bibinfo  {journal} {Am. Mineral.}\
  }\textbf {\bibinfo {volume} {80}},\ \bibinfo {pages} {329} (\bibinfo {year}
  {1995})}\BibitemShut {NoStop}%
\bibitem [{\citenamefont {Rulis}\ \emph {et~al.}(2004)\citenamefont {Rulis},
  \citenamefont {Ouyang},\ and\ \citenamefont {Ching}}]{Rulis04}%
  \BibitemOpen
  \bibfield  {author} {\bibinfo {author} {\bibfnamefont {P.}~\bibnamefont
  {Rulis}}, \bibinfo {author} {\bibfnamefont {L.}~\bibnamefont {Ouyang}}, \
  and\ \bibinfo {author} {\bibfnamefont {W.~Y.}\ \bibnamefont {Ching}},\
  }\href@noop {} {\bibfield  {journal} {\bibinfo  {journal} {Phys. Rev. B}\
  }\textbf {\bibinfo {volume} {70}},\ \bibinfo {pages} {155104} (\bibinfo
  {year} {2004})}\BibitemShut {NoStop}%
\bibitem [{\citenamefont {Sudarsan}\ \emph {et~al.}(1972)\citenamefont
  {Sudarsan}, \citenamefont {Young},\ and\ \citenamefont
  {Mackie}}]{sudarsan72}%
  \BibitemOpen
  \bibfield  {author} {\bibinfo {author} {\bibfnamefont {K.}~\bibnamefont
  {Sudarsan}}, \bibinfo {author} {\bibfnamefont {R.}~\bibnamefont {Young}}, \
  and\ \bibinfo {author} {\bibfnamefont {P.}~\bibnamefont {Mackie}},\
  }\href@noop {} {\bibfield  {journal} {\bibinfo  {journal} {Mater. Res.
  Bull.}\ }\textbf {\bibinfo {volume} {7}},\ \bibinfo {pages} {1331} (\bibinfo
  {year} {1972})}\BibitemShut {NoStop}%
\bibitem [{\citenamefont {Calzolari}\ and\ \citenamefont {{Buongiorno
  Nardelli}}(2013)}]{calzolari13}%
  \BibitemOpen
  \bibfield  {author} {\bibinfo {author} {\bibfnamefont {A.}~\bibnamefont
  {Calzolari}}\ and\ \bibinfo {author} {\bibfnamefont {M.}~\bibnamefont
  {{Buongiorno Nardelli}}},\ }\href@noop {} {\bibfield  {journal} {\bibinfo
  {journal} {Sci. Rep.}\ }\textbf {\bibinfo {volume} {3}},\ \bibinfo {pages}
  {2999} (\bibinfo {year} {2013})}\BibitemShut {NoStop}%
\bibitem [{\citenamefont {Tsuda}\ and\ \citenamefont {Arends}(1994)}]{tsuda94}%
  \BibitemOpen
  \bibfield  {author} {\bibinfo {author} {\bibfnamefont {H.}~\bibnamefont
  {Tsuda}}\ and\ \bibinfo {author} {\bibfnamefont {J.}~\bibnamefont {Arends}},\
  }\href@noop {} {\bibfield  {journal} {\bibinfo  {journal} {J. Dent. Res.}\
  }\textbf {\bibinfo {volume} {73}},\ \bibinfo {pages} {1703} (\bibinfo {year}
  {1994})}\BibitemShut {NoStop}%
\bibitem [{\citenamefont {Rehman}\ and\ \citenamefont
  {Bonfield}(1997)}]{rehman97}%
  \BibitemOpen
  \bibfield  {author} {\bibinfo {author} {\bibfnamefont {I.}~\bibnamefont
  {Rehman}}\ and\ \bibinfo {author} {\bibfnamefont {W.}~\bibnamefont
  {Bonfield}},\ }\href@noop {} {\bibfield  {journal} {\bibinfo  {journal} {J.
  Mater. Sci.: Mater Med.}\ }\textbf {\bibinfo {volume} {8}},\ \bibinfo {pages}
  {1} (\bibinfo {year} {1997})}\BibitemShut {NoStop}%
\bibitem [{\citenamefont {Cusc\'o}\ \emph {et~al.}(1998)\citenamefont
  {Cusc\'o}, \citenamefont {Guiti\'an}, \citenamefont {{de Aza}},\ and\
  \citenamefont {Art\'us}}]{cusco98}%
  \BibitemOpen
  \bibfield  {author} {\bibinfo {author} {\bibfnamefont {R.}~\bibnamefont
  {Cusc\'o}}, \bibinfo {author} {\bibfnamefont {F.}~\bibnamefont {Guiti\'an}},
  \bibinfo {author} {\bibfnamefont {S.}~\bibnamefont {{de Aza}}}, \ and\
  \bibinfo {author} {\bibfnamefont {L.}~\bibnamefont {Art\'us}},\ }\href@noop
  {} {\bibfield  {journal} {\bibinfo  {journal} {J. Eur. Chem. Soc.}\ }\textbf
  {\bibinfo {volume} {18}},\ \bibinfo {pages} {1301} (\bibinfo {year}
  {1998})}\BibitemShut {NoStop}%
\bibitem [{\citenamefont {Leroy}\ \emph {et~al.}(2000)\citenamefont {Leroy},
  \citenamefont {Leroy}, \citenamefont {Penel}, \citenamefont {Rey},
  \citenamefont {Lafforgue},\ and\ \citenamefont {Bres}}]{leroy}%
  \BibitemOpen
  \bibfield  {author} {\bibinfo {author} {\bibfnamefont {G.}~\bibnamefont
  {Leroy}}, \bibinfo {author} {\bibfnamefont {N.}~\bibnamefont {Leroy}},
  \bibinfo {author} {\bibfnamefont {G.}~\bibnamefont {Penel}}, \bibinfo
  {author} {\bibfnamefont {C.}~\bibnamefont {Rey}}, \bibinfo {author}
  {\bibfnamefont {P.}~\bibnamefont {Lafforgue}}, \ and\ \bibinfo {author}
  {\bibfnamefont {E.}~\bibnamefont {Bres}},\ }\href@noop {} {\bibfield
  {journal} {\bibinfo  {journal} {Appl. Spectr.}\ }\textbf {\bibinfo {volume}
  {54}},\ \bibinfo {pages} {1521} (\bibinfo {year} {2000})}\BibitemShut
  {NoStop}%
\bibitem [{\citenamefont {Klee}(1970)}]{klee}%
  \BibitemOpen
  \bibfield  {author} {\bibinfo {author} {\bibfnamefont {W.~E.}\ \bibnamefont
  {Klee}},\ }\href@noop {} {\bibfield  {journal} {\bibinfo  {journal} {Z.
  Kristallogr.}\ }\textbf {\bibinfo {volume} {131}},\ \bibinfo {pages} {95}
  (\bibinfo {year} {1970})}\BibitemShut {NoStop}%
\bibitem [{\citenamefont {Klee}\ and\ \citenamefont {Engel}(1970)}]{klee70}%
  \BibitemOpen
  \bibfield  {author} {\bibinfo {author} {\bibfnamefont {W.~E.}\ \bibnamefont
  {Klee}}\ and\ \bibinfo {author} {\bibfnamefont {G.}~\bibnamefont {Engel}},\
  }\href@noop {} {\bibfield  {journal} {\bibinfo  {journal} {J. Inorg. Nucl.
  Chem.}\ }\textbf {\bibinfo {volume} {32}},\ \bibinfo {pages} {1837} (\bibinfo
  {year} {1970})}\BibitemShut {NoStop}%
\bibitem [{\citenamefont {Penel}\ \emph {et~al.}(1997)\citenamefont {Penel},
  \citenamefont {Leroy}, \citenamefont {C.Rey}, \citenamefont {Sombret},
  \citenamefont {Huvenne},\ and\ \citenamefont {Bres}}]{penel97}%
  \BibitemOpen
  \bibfield  {author} {\bibinfo {author} {\bibfnamefont {G.}~\bibnamefont
  {Penel}}, \bibinfo {author} {\bibfnamefont {G.}~\bibnamefont {Leroy}},
  \bibinfo {author} {\bibnamefont {C.Rey}}, \bibinfo {author} {\bibfnamefont
  {B.}~\bibnamefont {Sombret}}, \bibinfo {author} {\bibfnamefont {J.~P.}\
  \bibnamefont {Huvenne}}, \ and\ \bibinfo {author} {\bibfnamefont
  {E.}~\bibnamefont {Bres}},\ }\href@noop {} {\bibfield  {journal} {\bibinfo
  {journal} {J. Mater. Sci: Mater. Med.}\ }\textbf {\bibinfo {volume} {8}},\
  \bibinfo {pages} {271} (\bibinfo {year} {1997})}\BibitemShut {NoStop}%
\bibitem [{\citenamefont {Baddiel}\ and\ \citenamefont
  {Berry}(1966)}]{berry66}%
  \BibitemOpen
  \bibfield  {author} {\bibinfo {author} {\bibfnamefont {C.~B.}\ \bibnamefont
  {Baddiel}}\ and\ \bibinfo {author} {\bibfnamefont {E.~E.}\ \bibnamefont
  {Berry}},\ }\href@noop {} {\bibfield  {journal} {\bibinfo  {journal} {Spectr.
  Acta}\ }\textbf {\bibinfo {volume} {22}},\ \bibinfo {pages} {1407} (\bibinfo
  {year} {1966})}\BibitemShut {NoStop}%
\bibitem [{\citenamefont {Wei}\ \emph {et~al.}(2003)\citenamefont {Wei},
  \citenamefont {Evans}, \citenamefont {Bostrom},\ and\ \citenamefont
  {Grondahl}}]{wei03}%
  \BibitemOpen
  \bibfield  {author} {\bibinfo {author} {\bibfnamefont {M.}~\bibnamefont
  {Wei}}, \bibinfo {author} {\bibfnamefont {J.~H.}\ \bibnamefont {Evans}},
  \bibinfo {author} {\bibfnamefont {T.}~\bibnamefont {Bostrom}}, \ and\
  \bibinfo {author} {\bibfnamefont {L.}~\bibnamefont {Grondahl}},\ }\href@noop
  {} {\bibfield  {journal} {\bibinfo  {journal} {J. Mater. Sci.: Mater Med.}\
  }\textbf {\bibinfo {volume} {14}},\ \bibinfo {pages} {311} (\bibinfo {year}
  {2003})}\BibitemShut {NoStop}%
\bibitem [{\citenamefont {Antonakos}\ \emph {et~al.}(2007)\citenamefont
  {Antonakos}, \citenamefont {Liarokapis},\ and\ \citenamefont
  {Leventouri}}]{antonakos07}%
  \BibitemOpen
  \bibfield  {author} {\bibinfo {author} {\bibfnamefont {A.}~\bibnamefont
  {Antonakos}}, \bibinfo {author} {\bibfnamefont {E.}~\bibnamefont
  {Liarokapis}}, \ and\ \bibinfo {author} {\bibfnamefont {T.}~\bibnamefont
  {Leventouri}},\ }\href@noop {} {\bibfield  {journal} {\bibinfo  {journal}
  {Biomater.}\ }\textbf {\bibinfo {volume} {28}},\ \bibinfo {pages} {3043}
  (\bibinfo {year} {2007})}\BibitemShut {NoStop}%
\bibitem [{\citenamefont {Giannozzi}\ \emph {et~al.}(2017)\citenamefont
  {Giannozzi}, \citenamefont {Andreussi}, \citenamefont {Brumme}, \citenamefont
  {Bunau}, \citenamefont {{Buongiorno Nardelli}}, \citenamefont {Calandra},
  \citenamefont {Car}, \citenamefont {Cavazzoni}, \citenamefont {Ceresoli},
  \citenamefont {Cococcioni}, \citenamefont {Colonna}, \citenamefont
  {Carnimeo}, \citenamefont {Corso}, \citenamefont {de~Gironcoli},
  \citenamefont {Delugas}, \citenamefont {DiStasio}, \citenamefont {Ferretti},
  \citenamefont {Floris}, \citenamefont {Fratesi}, \citenamefont {Fugallo},
  \citenamefont {Gebauer}, \citenamefont {Gerstmann}, \citenamefont {Giustino},
  \citenamefont {Gorni}, \citenamefont {Jia}, \citenamefont {Kawamura},
  \citenamefont {Ko}, \citenamefont {Kokalj}, \citenamefont
  {K\"{u}\c{c}kbenli}, \citenamefont {Lazzeri}, \citenamefont {Marsili},
  \citenamefont {Marzari}, \citenamefont {Mauri}, \citenamefont {Nguyen},
  \citenamefont {Nguyen}, \citenamefont {de-la Roza}, \citenamefont {Paulatto},
  \citenamefont {Ponc{\'{e}}}, \citenamefont {Rocca}, \citenamefont {Sabatini},
  \citenamefont {Santra}, \citenamefont {Schlipf}, \citenamefont {Seitsonen},
  \citenamefont {Smogunov}, \citenamefont {Timrov}, \citenamefont {Thonhauser},
  \citenamefont {Umari}, \citenamefont {Vast}, \citenamefont {Wu},\ and\
  \citenamefont {Baroni}}]{giannozzi17}%
  \BibitemOpen
  \bibfield  {author} {\bibinfo {author} {\bibfnamefont {P.}~\bibnamefont
  {Giannozzi}}, \bibinfo {author} {\bibfnamefont {O.}~\bibnamefont
  {Andreussi}}, \bibinfo {author} {\bibfnamefont {T.}~\bibnamefont {Brumme}},
  \bibinfo {author} {\bibfnamefont {O.}~\bibnamefont {Bunau}}, \bibinfo
  {author} {\bibfnamefont {M.}~\bibnamefont {{Buongiorno Nardelli}}}, \bibinfo
  {author} {\bibfnamefont {M.}~\bibnamefont {Calandra}}, \bibinfo {author}
  {\bibfnamefont {R.}~\bibnamefont {Car}}, \bibinfo {author} {\bibfnamefont
  {C.}~\bibnamefont {Cavazzoni}}, \bibinfo {author} {\bibfnamefont
  {D.}~\bibnamefont {Ceresoli}}, \bibinfo {author} {\bibfnamefont
  {M.}~\bibnamefont {Cococcioni}}, \bibinfo {author} {\bibfnamefont
  {N.}~\bibnamefont {Colonna}}, \bibinfo {author} {\bibfnamefont
  {I.}~\bibnamefont {Carnimeo}}, \bibinfo {author} {\bibfnamefont {A.~D.}\
  \bibnamefont {Corso}}, \bibinfo {author} {\bibfnamefont {S.}~\bibnamefont
  {de~Gironcoli}}, \bibinfo {author} {\bibfnamefont {P.}~\bibnamefont
  {Delugas}}, \bibinfo {author} {\bibfnamefont {R.~A.}\ \bibnamefont
  {DiStasio}}, \bibinfo {author} {\bibfnamefont {A.}~\bibnamefont {Ferretti}},
  \bibinfo {author} {\bibfnamefont {A.}~\bibnamefont {Floris}}, \bibinfo
  {author} {\bibfnamefont {G.}~\bibnamefont {Fratesi}}, \bibinfo {author}
  {\bibfnamefont {G.}~\bibnamefont {Fugallo}}, \bibinfo {author} {\bibfnamefont
  {R.}~\bibnamefont {Gebauer}}, \bibinfo {author} {\bibfnamefont
  {U.}~\bibnamefont {Gerstmann}}, \bibinfo {author} {\bibfnamefont
  {F.}~\bibnamefont {Giustino}}, \bibinfo {author} {\bibfnamefont
  {T.}~\bibnamefont {Gorni}}, \bibinfo {author} {\bibfnamefont
  {J.}~\bibnamefont {Jia}}, \bibinfo {author} {\bibfnamefont {M.}~\bibnamefont
  {Kawamura}}, \bibinfo {author} {\bibfnamefont {H.-Y.}\ \bibnamefont {Ko}},
  \bibinfo {author} {\bibfnamefont {A.}~\bibnamefont {Kokalj}}, \bibinfo
  {author} {\bibfnamefont {E.}~\bibnamefont {K\"{u}\c{c}kbenli}}, \bibinfo
  {author} {\bibfnamefont {M.}~\bibnamefont {Lazzeri}}, \bibinfo {author}
  {\bibfnamefont {M.}~\bibnamefont {Marsili}}, \bibinfo {author} {\bibfnamefont
  {N.}~\bibnamefont {Marzari}}, \bibinfo {author} {\bibfnamefont
  {F.}~\bibnamefont {Mauri}}, \bibinfo {author} {\bibfnamefont {N.~L.}\
  \bibnamefont {Nguyen}}, \bibinfo {author} {\bibfnamefont {H.-V.}\
  \bibnamefont {Nguyen}}, \bibinfo {author} {\bibfnamefont {A.~O.}\
  \bibnamefont {de-la Roza}}, \bibinfo {author} {\bibfnamefont
  {L.}~\bibnamefont {Paulatto}}, \bibinfo {author} {\bibfnamefont
  {S.}~\bibnamefont {Ponc{\'{e}}}}, \bibinfo {author} {\bibfnamefont
  {D.}~\bibnamefont {Rocca}}, \bibinfo {author} {\bibfnamefont
  {R.}~\bibnamefont {Sabatini}}, \bibinfo {author} {\bibfnamefont
  {B.}~\bibnamefont {Santra}}, \bibinfo {author} {\bibfnamefont
  {M.}~\bibnamefont {Schlipf}}, \bibinfo {author} {\bibfnamefont {A.~P.}\
  \bibnamefont {Seitsonen}}, \bibinfo {author} {\bibfnamefont {A.}~\bibnamefont
  {Smogunov}}, \bibinfo {author} {\bibfnamefont {I.}~\bibnamefont {Timrov}},
  \bibinfo {author} {\bibfnamefont {T.}~\bibnamefont {Thonhauser}}, \bibinfo
  {author} {\bibfnamefont {P.}~\bibnamefont {Umari}}, \bibinfo {author}
  {\bibfnamefont {N.}~\bibnamefont {Vast}}, \bibinfo {author} {\bibfnamefont
  {X.}~\bibnamefont {Wu}}, \ and\ \bibinfo {author} {\bibfnamefont
  {S.}~\bibnamefont {Baroni}},\ }\href@noop {} {\bibfield  {journal} {\bibinfo
  {journal} {J. Phys.: Cond. Matt.}\ }\textbf {\bibinfo {volume} {29}},\
  \bibinfo {pages} {465901} (\bibinfo {year} {2017})}\BibitemShut {NoStop}%
\bibitem [{\citenamefont {Thonhauser}\ \emph {et~al.}(2015)\citenamefont
  {Thonhauser}, \citenamefont {Zuluaga}, \citenamefont {Arter}, \citenamefont
  {Berland}, \citenamefont {Schr\"oder},\ and\ \citenamefont
  {Hyldgaard}}]{Thonhauser15}%
  \BibitemOpen
  \bibfield  {author} {\bibinfo {author} {\bibfnamefont {T.}~\bibnamefont
  {Thonhauser}}, \bibinfo {author} {\bibfnamefont {S.}~\bibnamefont {Zuluaga}},
  \bibinfo {author} {\bibfnamefont {C.~A.}\ \bibnamefont {Arter}}, \bibinfo
  {author} {\bibfnamefont {K.}~\bibnamefont {Berland}}, \bibinfo {author}
  {\bibfnamefont {E.}~\bibnamefont {Schr\"oder}}, \ and\ \bibinfo {author}
  {\bibfnamefont {P.}~\bibnamefont {Hyldgaard}},\ }\href@noop {} {\bibfield
  {journal} {\bibinfo  {journal} {Phys. Rev. Lett.}\ }\textbf {\bibinfo
  {volume} {115}},\ \bibinfo {pages} {136402} (\bibinfo {year}
  {2015})}\BibitemShut {NoStop}%
\bibitem [{\citenamefont {Langreth}\ \emph {et~al.}(2009)\citenamefont
  {Langreth}, \citenamefont {Lundqvist}, \citenamefont {Chakarova-K\"ack},
  \citenamefont {Cooper}, \citenamefont {Dion}, \citenamefont {Hyldgaard},
  \citenamefont {Kelkkanen}, \citenamefont {Kleis}, \citenamefont {Kong},
  \citenamefont {Li}, \citenamefont {Moses}, \citenamefont {Murray},
  \citenamefont {Puzder}, \citenamefont {Rydberg}, \citenamefont {Schr\"oder},\
  and\ \citenamefont {Thonhauser}}]{Langreth15}%
  \BibitemOpen
  \bibfield  {author} {\bibinfo {author} {\bibfnamefont {D.~C.}\ \bibnamefont
  {Langreth}}, \bibinfo {author} {\bibfnamefont {B.~I.}\ \bibnamefont
  {Lundqvist}}, \bibinfo {author} {\bibfnamefont {S.~D.}\ \bibnamefont
  {Chakarova-K\"ack}}, \bibinfo {author} {\bibfnamefont {V.~R.}\ \bibnamefont
  {Cooper}}, \bibinfo {author} {\bibfnamefont {M.}~\bibnamefont {Dion}},
  \bibinfo {author} {\bibfnamefont {P.}~\bibnamefont {Hyldgaard}}, \bibinfo
  {author} {\bibfnamefont {A.}~\bibnamefont {Kelkkanen}}, \bibinfo {author}
  {\bibfnamefont {J.}~\bibnamefont {Kleis}}, \bibinfo {author} {\bibfnamefont
  {L.}~\bibnamefont {Kong}}, \bibinfo {author} {\bibfnamefont {S.}~\bibnamefont
  {Li}}, \bibinfo {author} {\bibfnamefont {P.~G.}\ \bibnamefont {Moses}},
  \bibinfo {author} {\bibfnamefont {E.}~\bibnamefont {Murray}}, \bibinfo
  {author} {\bibfnamefont {A.}~\bibnamefont {Puzder}}, \bibinfo {author}
  {\bibfnamefont {H.}~\bibnamefont {Rydberg}}, \bibinfo {author} {\bibfnamefont
  {E.}~\bibnamefont {Schr\"oder}}, \ and\ \bibinfo {author} {\bibfnamefont
  {T.}~\bibnamefont {Thonhauser}},\ }\href@noop {} {\bibfield  {journal}
  {\bibinfo  {journal} {J. Phys.: Conden. Matt.}\ }\textbf {\bibinfo {volume}
  {21}},\ \bibinfo {pages} {084203} (\bibinfo {year} {2009})}\BibitemShut
  {NoStop}%
\bibitem [{\citenamefont {Vanderbilt}(1990)}]{vanderbilt90}%
  \BibitemOpen
  \bibfield  {author} {\bibinfo {author} {\bibfnamefont {D.}~\bibnamefont
  {Vanderbilt}},\ }\href@noop {} {\bibfield  {journal} {\bibinfo  {journal}
  {Phys.~Rev.~B}\ }\textbf {\bibinfo {volume} {41}},\ \bibinfo {pages} {R7892}
  (\bibinfo {year} {1990})}\BibitemShut {NoStop}%
\bibitem [{\citenamefont {Korotin}\ \emph {et~al.}(2000)\citenamefont
  {Korotin}, \citenamefont {Fujiwara},\ and\ \citenamefont
  {Anisimov}}]{Anisimov00}%
  \BibitemOpen
  \bibfield  {author} {\bibinfo {author} {\bibfnamefont {M.}~\bibnamefont
  {Korotin}}, \bibinfo {author} {\bibfnamefont {T.}~\bibnamefont {Fujiwara}}, \
  and\ \bibinfo {author} {\bibfnamefont {V.}~\bibnamefont {Anisimov}},\
  }\href@noop {} {\bibfield  {journal} {\bibinfo  {journal} {Phys. Rev. B}\
  }\textbf {\bibinfo {volume} {62}},\ \bibinfo {pages} {5696} (\bibinfo {year}
  {2000})}\BibitemShut {NoStop}%
\bibitem [{\citenamefont {Cococcioni}\ and\ \citenamefont {{de
  Gironcoli}}(2005)}]{Cococcioni05}%
  \BibitemOpen
  \bibfield  {author} {\bibinfo {author} {\bibfnamefont {M.}~\bibnamefont
  {Cococcioni}}\ and\ \bibinfo {author} {\bibfnamefont {S.}~\bibnamefont {{de
  Gironcoli}}},\ }\href@noop {} {\bibfield  {journal} {\bibinfo  {journal}
  {Phys. Rev. B}\ }\textbf {\bibinfo {volume} {71}},\ \bibinfo {pages} {035105}
  (\bibinfo {year} {2005})}\BibitemShut {NoStop}%
\bibitem [{\citenamefont {Agapito}\ \emph {et~al.}(2015)\citenamefont
  {Agapito}, \citenamefont {Curtarolo},\ and\ \citenamefont {{Buongiorno
  Nardelli}}}]{Agapito2015}%
  \BibitemOpen
  \bibfield  {author} {\bibinfo {author} {\bibfnamefont {L.~A.}\ \bibnamefont
  {Agapito}}, \bibinfo {author} {\bibfnamefont {S.}~\bibnamefont {Curtarolo}},
  \ and\ \bibinfo {author} {\bibfnamefont {M.}~\bibnamefont {{Buongiorno
  Nardelli}}},\ }\href@noop {} {\bibfield  {journal} {\bibinfo  {journal}
  {Phys. Rev. X}\ }\textbf {\bibinfo {volume} {5}},\ \bibinfo {pages} {1}
  (\bibinfo {year} {2015})}\BibitemShut {NoStop}%
\bibitem [{\citenamefont {Gopal}\ \emph {et~al.}(2015)\citenamefont {Gopal},
  \citenamefont {Fornari}, \citenamefont {Curtarolo}, \citenamefont {Agapito},
  \citenamefont {Liyanage},\ and\ \citenamefont {{Buongiorno
  Nardelli}}}]{Gopal:2015bf}%
  \BibitemOpen
  \bibfield  {author} {\bibinfo {author} {\bibfnamefont {P.}~\bibnamefont
  {Gopal}}, \bibinfo {author} {\bibfnamefont {M.}~\bibnamefont {Fornari}},
  \bibinfo {author} {\bibfnamefont {S.}~\bibnamefont {Curtarolo}}, \bibinfo
  {author} {\bibfnamefont {L.~A.}\ \bibnamefont {Agapito}}, \bibinfo {author}
  {\bibfnamefont {L.~S.~I.}\ \bibnamefont {Liyanage}}, \ and\ \bibinfo {author}
  {\bibfnamefont {M.}~\bibnamefont {{Buongiorno Nardelli}}},\ }\href@noop {}
  {\bibfield  {journal} {\bibinfo  {journal} {Phys. Rev. B}\ }\textbf {\bibinfo
  {volume} {91}},\ \bibinfo {pages} {245202} (\bibinfo {year}
  {2015})}\BibitemShut {NoStop}%
\bibitem [{\citenamefont {Eaton}\ \emph {et~al.}(2018)\citenamefont {Eaton},
  \citenamefont {Catellani},\ and\ \citenamefont {Calzolari}}]{eaton18}%
  \BibitemOpen
  \bibfield  {author} {\bibinfo {author} {\bibfnamefont {M.}~\bibnamefont
  {Eaton}}, \bibinfo {author} {\bibfnamefont {A.}~\bibnamefont {Catellani}}, \
  and\ \bibinfo {author} {\bibfnamefont {A.}~\bibnamefont {Calzolari}},\
  }\href@noop {} {\bibfield  {journal} {\bibinfo  {journal} {Opt. Express}\
  }\textbf {\bibinfo {volume} {26}},\ \bibinfo {pages} {5342 (1} (\bibinfo
  {year} {2018})}\BibitemShut {NoStop}%
\bibitem [{\citenamefont {Cigarini}\ \emph {et~al.}(2018)\citenamefont
  {Cigarini}, \citenamefont {Ruini}, \citenamefont {Catellani},\ and\
  \citenamefont {Calzolari}}]{cigarini18}%
  \BibitemOpen
  \bibfield  {author} {\bibinfo {author} {\bibfnamefont {L.}~\bibnamefont
  {Cigarini}}, \bibinfo {author} {\bibfnamefont {A.}~\bibnamefont {Ruini}},
  \bibinfo {author} {\bibfnamefont {A.}~\bibnamefont {Catellani}}, \ and\
  \bibinfo {author} {\bibfnamefont {A.}~\bibnamefont {Calzolari}},\ }\href@noop
  {} {\bibfield  {journal} {\bibinfo  {journal} {Phys. Chem. Chem. Phys.}\
  }\textbf {\bibinfo {volume} {20}},\ \bibinfo {pages} {5021} (\bibinfo {year}
  {2018})}\BibitemShut {NoStop}%
\bibitem [{\citenamefont {Supka}\ \emph {et~al.}(2017)\citenamefont {Supka},
  \citenamefont {Lyons}, \citenamefont {Liyanage}, \citenamefont {{D'Amico}},
  \citenamefont {{Al Orabi}}, \citenamefont {Mahatara}, \citenamefont {Gopal},
  \citenamefont {Toher}, \citenamefont {Ceresoli}, \citenamefont {Calzolari},
  \citenamefont {Curtarolo}, \citenamefont {{Buongiorno Nardelli}},\ and\
  \citenamefont {Fornari}}]{aflowpi}%
  \BibitemOpen
  \bibfield  {author} {\bibinfo {author} {\bibfnamefont {A.~R.}\ \bibnamefont
  {Supka}}, \bibinfo {author} {\bibfnamefont {T.~E.}\ \bibnamefont {Lyons}},
  \bibinfo {author} {\bibfnamefont {L.}~\bibnamefont {Liyanage}}, \bibinfo
  {author} {\bibfnamefont {P.}~\bibnamefont {{D'Amico}}}, \bibinfo {author}
  {\bibfnamefont {R.~A.~R.}\ \bibnamefont {{Al Orabi}}}, \bibinfo {author}
  {\bibfnamefont {S.}~\bibnamefont {Mahatara}}, \bibinfo {author}
  {\bibfnamefont {P.}~\bibnamefont {Gopal}}, \bibinfo {author} {\bibfnamefont
  {C.}~\bibnamefont {Toher}}, \bibinfo {author} {\bibfnamefont
  {D.}~\bibnamefont {Ceresoli}}, \bibinfo {author} {\bibfnamefont
  {A.}~\bibnamefont {Calzolari}}, \bibinfo {author} {\bibfnamefont
  {S.}~\bibnamefont {Curtarolo}}, \bibinfo {author} {\bibfnamefont
  {M.}~\bibnamefont {{Buongiorno Nardelli}}}, \ and\ \bibinfo {author}
  {\bibfnamefont {M.}~\bibnamefont {Fornari}},\ }\href@noop {} {\bibfield
  {journal} {\bibinfo  {journal} {Comput. Mater. Sci.}\ }\textbf {\bibinfo
  {volume} {136}},\ \bibinfo {pages} {76} (\bibinfo {year} {2017})}\BibitemShut
  {NoStop}%
\bibitem [{\citenamefont {{Buongiorno Nardelli}}\ \emph
  {et~al.}(2018)\citenamefont {{Buongiorno Nardelli}}, \citenamefont
  {Cerasoli}, \citenamefont {Costa}, \citenamefont {Curtarolo}, \citenamefont
  {Gennaro}, \citenamefont {Fornari}, \citenamefont {Liyanage}, \citenamefont
  {Supka},\ and\ \citenamefont {Wang}}]{paoflow}%
  \BibitemOpen
  \bibfield  {author} {\bibinfo {author} {\bibfnamefont {M.}~\bibnamefont
  {{Buongiorno Nardelli}}}, \bibinfo {author} {\bibfnamefont {F.~T.}\
  \bibnamefont {Cerasoli}}, \bibinfo {author} {\bibfnamefont {M.}~\bibnamefont
  {Costa}}, \bibinfo {author} {\bibfnamefont {S.}~\bibnamefont {Curtarolo}},
  \bibinfo {author} {\bibfnamefont {R.~D.}\ \bibnamefont {Gennaro}}, \bibinfo
  {author} {\bibfnamefont {M.}~\bibnamefont {Fornari}}, \bibinfo {author}
  {\bibfnamefont {L.}~\bibnamefont {Liyanage}}, \bibinfo {author}
  {\bibfnamefont {A.~R.}\ \bibnamefont {Supka}}, \ and\ \bibinfo {author}
  {\bibfnamefont {H.}~\bibnamefont {Wang}},\ }\href@noop {} {\bibfield
  {journal} {\bibinfo  {journal} {Comput. Mater. Sci.}\ }\textbf {\bibinfo
  {volume} {143}},\ \bibinfo {pages} {462} (\bibinfo {year}
  {2018})}\BibitemShut {NoStop}%
\bibitem [{\citenamefont {Tomasi}\ \emph {et~al.}(2005)\citenamefont {Tomasi},
  \citenamefont {Mennucci},\ and\ \citenamefont {Cammi}}]{tomasi}%
  \BibitemOpen
  \bibfield  {author} {\bibinfo {author} {\bibfnamefont {J.}~\bibnamefont
  {Tomasi}}, \bibinfo {author} {\bibfnamefont {B.}~\bibnamefont {Mennucci}}, \
  and\ \bibinfo {author} {\bibfnamefont {R.}~\bibnamefont {Cammi}},\
  }\href@noop {} {\bibfield  {journal} {\bibinfo  {journal} {Chem. Rev.}\
  }\textbf {\bibinfo {volume} {105}},\ \bibinfo {pages} {2999} (\bibinfo {year}
  {2005})}\BibitemShut {NoStop}%
\bibitem [{\citenamefont {Andreussi}\ \emph {et~al.}(2012)\citenamefont
  {Andreussi}, \citenamefont {Dabo},\ and\ \citenamefont
  {Marzari}}]{andreussi}%
  \BibitemOpen
  \bibfield  {author} {\bibinfo {author} {\bibfnamefont {O.}~\bibnamefont
  {Andreussi}}, \bibinfo {author} {\bibfnamefont {I.}~\bibnamefont {Dabo}}, \
  and\ \bibinfo {author} {\bibfnamefont {N.}~\bibnamefont {Marzari}},\
  }\href@noop {} {\bibfield  {journal} {\bibinfo  {journal} {J. Chem. Phys.}\
  }\textbf {\bibinfo {volume} {136}},\ \bibinfo {pages} {064102} (\bibinfo
  {year} {2012})}\BibitemShut {NoStop}%
\end{thebibliography}%

\section*{ACKNOWLEDGEMENTS}
\noindent 
The authors thank Dr. Andrew Supka for technical support and Dr. Mary J. Tecklenburg for suggesting the problem.
M.F. and M.B.N. acknowledge collaboration with the AFLOW Consortium (http://www.aflow.org) under the sponsorship of DOD-ONR (Grants N000141310635 and N000141512266).

\section*{AUTHOR CONTRIBUTIONS STATEMENT}
\noindent 
{\small A.C. coordinated the project providing calculations, data analysis and co-writing the text; B.P. generated the initial structures; S.C. co-wrote the manuscript; M.B.N. analysed the results and co-wrote the manuscript; M.F. conceived the project and co-wrote the manuscript.}

\section*{ADDITIONAL INFORMATION}
{\small 
\subsection*{Supplementary information}
\noindent Supplementary Information includes details on  the crystalline structure of hexagonal apatite crystals (Sec. S1.); the method (Tab 1) and on the electronic structure (Sec. S2)
the X-ray spectra (Sec. S3) and  the dielectric and vibrational properties (Sec. S4) of FAp, HAp and CFAp systems; the  polarized (Sec. S5)  and isotope-substituted
(Sec. S6) Raman spectra as well as the polarized Raman spectra of alternative apatite crystals (Sec. S7).

\subsection*{Competing interests} 
\noindent The authors declare no competing interests.}

\newpage
\section*{Figure Legends}

\newpage
\section*{Tables}

\begin{table}[!h!]
\caption{\small {\bf Solvated orthophophate.} 
Vibrational frequencies (cm$^{-1}$) of isolated PO$_4^{3-}$.
}
\begin{center}
\begin{tabular}{ccccc}
\hline
 	 	&  $\nu_2$	& $\nu_4$		& $\nu_1$		& $\nu_3$ \\
\hline \hline
present	&  419    		& 587		& 951		& 1032      \\
(water)	&			&			&			&		\\
present	&  412    		& 588		& 931		& 1013      \\
(apatite)	&			&			&			&		\\
exp$^a$	& 420		& 567		& 938		& 1017 \\
(water)	&			&			&			&		\\
\hline
   \end{tabular}\\
$^a$ Ref. \citenum{schultze73}
\end{center}\label{tab:po4}
\end{table}

\begin{table}[!h!]
\caption{\small {\bf PO$_4^{3-}$ modes in apatite spectra.} Band position (cm$^{-1}$) corresponding to orthophosphate group in apatites. Experimental data include both IR and Raman measurements.}
\begin{center}
\begin{tabular}{cccccc}
\hline
	&mode 	 	&  Present		&	Exp$^a$	& Exp$^b$	& Exp$^c$ \\
\hline
FAp	& $\nu_2$		& 437--460	& 315--471	& 429--470	& 431--473         \\
	& $\nu_4$		& 591--654	& 566--603	& 560--615	& 569--6151\\
	& $\nu_1$		& 1011		& 960 		& 962		& 963        \\
	& $\nu_3$		& 1065--1165	& 1040--1200	& 1032--1090	& 1034-1104 \\
        &			&			&			&			& \\
\hline
	&mode 	 	&  Present		&	Exp$^a$	& Exp$^d$	& Exp$^e$ \\
	\hline
HAp	& $\nu_2$		& 445--484	& 350--472	&433--448		& 473 	\\
	& $\nu_4$		& 595--661	& 564--725	& 580--614	& 570--601 \\
	& $\nu_1$		& 1013 		&950--980		& 963		& 962 	\\
	& $\nu_3$		& 1023--1194	& 1028-1209	& 1029-1077	& 1042-1088 \\
        &			&			&			&			& \\
\hline
	&mode 	 	&  Present		&	Exp$^f$	&  			&   \\
	\hline
CFAp& $\nu_2$		&   434--468	& 440-557   	&			&  \\
 	& $\nu_4$		&   590--655	& 570--602	&			&  \\
	& $\nu_1$		& 994		& 966		&			&  \\
	& $\nu_3$		& 1003--1178	& 1083--1156	&			& \\
\hline
\end{tabular}\\
$^a$ Ref. \citenum{berry66},
$^b$ Ref. \citenum{klee},
$^c$ Ref. \citenum{leroy},
$^d$ Ref. \citenum{penel97},
$^e$ Ref. 	\citenum{klee70},
$^f$	Ref. \citenum{antonakos07}
\end{center}\label{tab:var}
\end{table}

\end{document}